\documentclass[10pt,conference]{IEEEtran}

\usepackage{graphicx}
\usepackage{amsmath}
\usepackage[hidelinks]{hyperref}

\usepackage{rotating}
\usepackage{tabularx}
\usepackage{array}
\usepackage{xurl}
\newcolumntype{Y}{>{\raggedright\arraybackslash}X}
\usepackage{dblfloatfix}
\usepackage{longtable}
\usepackage{booktabs}
\usepackage{placeins}

\usepackage{tikz}
\usetikzlibrary{positioning,arrows.meta,shapes.geometric,shapes.symbols,calc}

\usepackage{listings}

\usepackage{booktabs}
\usepackage[caption=false,font=footnotesize]{subfig}

\usepackage{makecell}

\usepackage{algorithm}
\usepackage{algpseudocode}

\usepackage{float} 

\title{A Ground-Truth-Based Evaluation of Vulnerability Detection Across Multiple Ecosystems}

\author{
\centering
\IEEEauthorblockN{
Peter Mandl\IEEEauthorrefmark{1},
Paul Mandl\IEEEauthorrefmark{2},
Martin H\"ausl\IEEEauthorrefmark{1},
Maximilian Auch\IEEEauthorrefmark{1}
}
\IEEEauthorblockA{\IEEEauthorrefmark{1}
University of Applied Sciences Munich, Munich, Germany\\
peter.mandl@hm.edu,
martin.haeusl@hm.edu,
maximilian.auch@hm.edu
}
\IEEEauthorblockA{\IEEEauthorrefmark{2}
Findustrial, Munich\\
paul.mandl@evaltech.de
}
}

\begin{document}

\maketitle
\begin{abstract}

Automated vulnerability detection tools are widely used to
identify security vulnerabilities in software dependencies.
However, the evaluation of such tools remains challenging due
to the heterogeneous structure of vulnerability data sources,
inconsistent identifier schemes, and ambiguities in version
range specifications.

In this paper, we present an empirical evaluation of vulnerability
detection across multiple software ecosystems using a curated
ground-truth dataset derived from the Open Source Vulnerabilities
(OSV) database. The dataset explicitly maps vulnerabilities to
concrete package versions and enables a systematic comparison of
detection results across different tools and services.

Since vulnerability databases such as OSV are continuously
updated, the dataset used in this study represents a snapshot
of the vulnerability landscape at the time of the evaluation.
To support reproducibility and future studies, we provide an
open-source tool that automatically reconstructs the dataset
from the current OSV database using the methodology described
in this paper.

Our evaluation highlights systematic differences between
vulnerability detection systems and demonstrates the importance
of transparent dataset construction for reproducible empirical
security research.

\end{abstract}

\noindent\textit{
This manuscript has not undergone peer review and may be revised.
}



\section{Introduction}

Software systems increasingly rely on large numbers of external
dependencies obtained from open-source package ecosystems.
While this reuse accelerates software development, it also
introduces security risks, since vulnerabilities in third-party
components may propagate into dependent applications. As a
result, automated vulnerability detection tools have become
widely used to identify vulnerable dependencies and support
secure software development practices.

Despite their widespread adoption, evaluating the effectiveness
of vulnerability detection systems remains challenging. One
major difficulty arises from the heterogeneous structure of
vulnerability information sources. Vulnerability data is
distributed across multiple databases and advisory systems,
which differ in their identifier schemes, metadata models,
and representations of affected software components.

In particular, vulnerability databases may describe affected
software either using product-centric identifiers such as
Common Platform Enumeration (CPE) entries or using ecosystem-
specific component identifiers. In addition, vulnerability
advisories frequently specify affected software using version
ranges whose interpretation depends on ecosystem-specific
versioning rules. These differences complicate the automated
interpretation of vulnerability information and may lead to
inconsistent detection results across tools.

A further challenge for empirical studies is the absence of a
stable and widely accepted ground-truth describing which
software component versions are affected by specific
vulnerabilities. Vulnerability databases such as the Open
Source Vulnerabilities (OSV) database continuously evolve as
new vulnerabilities are disclosed and existing records are
updated. Consequently, any evaluation dataset derived from
such sources represents only a snapshot of the vulnerability
landscape at a particular point in time.

To address these challenges, this paper introduces a curated
ground-truth dataset derived from the OSV database that
explicitly maps vulnerabilities to concrete package versions
across multiple software ecosystems. The dataset enables a
systematic comparison of vulnerability detection results
across different tools and services.

Because vulnerability databases evolve continuously, the
dataset used in this study represents a snapshot of the OSV
database at the time of the evaluation. To support
reproducibility and enable future studies, we developed an
open-source tool that automatically reconstructs the dataset
from the current OSV database using the dataset construction
process described in this paper.

Using this dataset, we perform a comparative evaluation of
representative vulnerability detection systems and analyze
systematic differences in their detection results.

The main contributions of this paper are:

\begin{itemize}
\item A methodology for constructing a ground-truth dataset
mapping vulnerabilities to concrete package versions across
multiple software ecosystems.

\item A curated dataset derived from the OSV database that
enables systematic comparison of vulnerability detection
results.

\item An empirical evaluation of representative vulnerability
detection tools and services using the constructed dataset.

\item An open-source tool that allows researchers to
reconstruct and update the dataset from current OSV data,
supporting reproducible vulnerability detection research.
\end{itemize}

The remainder of this paper is structured as follows. Section~II
introduces the relevant background on vulnerability databases,
identifier schemes, and version semantics. Section~III describes
the construction of the ground-truth dataset. Section~IV presents
the evaluation methodology. Section~V introduces the analyzed
vulnerability detection tools and their data sources. Section~VI
reports the evaluation results. Section~VII discusses the
findings and limitations. Section~VIII situates the work in
relation to prior research, and Section~IX concludes the paper
and outlines directions for future work.

\section{Background}

This section introduces the vulnerability data sources, identifier systems, and
version semantics that underpin our evaluation. We further clarify the
conceptual distinction between vulnerabilities and advisories, which is central
to understanding the limitations of automated vulnerability detection tools.

\subsection{Vulnerability Identifiers and Data Sources}

Vulnerability detection systems rely on publicly available
vulnerability information provided by several heterogeneous
data sources. A central role is played by the Common
Vulnerabilities and Exposures (CVE) system \cite{cve}, which
assigns identifiers to publicly disclosed vulnerabilities.
CVE identifiers serve as a common reference across vulnerability
databases, tools, and advisory systems.

A major enrichment source for CVE entries is the National
Vulnerability Database (NVD) \cite{nvd}. The NVD extends CVE
records with additional metadata, including severity
information, weakness classifications, and affected platform
information. For the latter, the NVD commonly relies on the
Common Platform Enumeration (CPE) scheme \cite{cpe}, which
provides a standardized naming format for software products
and versions.

In addition to CVE/NVD-based reporting, ecosystem-specific
advisory systems have become increasingly important.
GitHub Security Advisories (GHSA) \cite{ghsa}, for example,
provide vulnerability information with an explicit focus on
open-source packages and affected component versions.
Similarly, the Open Source Vulnerabilities (OSV) database
\cite{osv} aggregates vulnerability information across
multiple ecosystems and represents vulnerabilities in terms
of ecosystem identifiers, package names, and affected
versions.

These sources differ considerably in scope, coverage, and
growth dynamics. Table~\ref{tab:vuln-databases} provides an
overview of the approximate size and recent growth of several
widely used vulnerability identifier catalogs and advisory
databases.

\begin{table*}[t]
\centering
\caption{Approximate size and recent growth per year of major vulnerability
identifier catalogs and advisory databases}
\label{tab:vuln-databases}
\small
\setlength{\tabcolsep}{5pt}
\renewcommand{\arraystretch}{1.1}

\begin{tabular*}{\textwidth}{@{\extracolsep{\fill}}
  >{\raggedright\arraybackslash}p{0.23\textwidth}
  >{\centering\arraybackslash}p{0.13\textwidth}
  >{\centering\arraybackslash}p{0.14\textwidth}
  >{\raggedright\arraybackslash}p{0.42\textwidth}
@{}}
\toprule
Vulnerability Source &
Approx.\ Size &
Growth (\(\approx\) 2025) per year &
Scope \\
\midrule

CVE (MITRE) &
$\sim3.3\times10^5$ &
$5$--$8\%$ &
Global catalog of publicly disclosed vulnerability identifiers
serving as the primary reference for vulnerability tracking \\

NVD (NIST) &
$\sim3.3\times10^5$ &
$5$--$8\%$ &
CVE records enriched with additional metadata such as CVSS
severity scores and CPE-based product information \\

GitHub Advisory Database (GHSA) &
$\sim2.5$--$3.0\times10^4$ &
$15$--$25\%$ &
Package-level security advisories curated by GitHub,
including both CVE and non-CVE disclosures \\

Sonatype OSS Index &
$\sim1.5\times10^4$ &
$8$--$12\%$ &
Curated vulnerability advisories mapped to package
coordinates across multiple ecosystems \\

OSV (aggregated vulnerability database) &
$\sim1\times10^5$ &
$20$--$30\%$ &
Aggregated vulnerability records across multiple ecosystems,
linking vulnerabilities to packages and affected versions \\

PyPI advisories &
$\sim1.7\times10^4$ &
$10$--$15\%$ &
Security advisories affecting Python packages \\

npm advisories &
$\sim2.1\times10^5$ &
$25$--$35\%$ &
Security advisories affecting JavaScript packages \\

Maven advisories &
$\sim6\times10^3$ &
$3$--$6\%$ &
Security advisories affecting Java artifacts \\

NuGet advisories &
$\sim1.5\times10^3$ &
$2$--$5\%$ &
Security advisories affecting .NET packages \\

\bottomrule
\end{tabular*}

\smallskip
\raggedright
\footnotesize
\textit{Note:} Growth figures denote approximate annual changes
observed around 2024--2025 and are reported as ranges to reflect
differences in counting methodologies across databases.
\end{table*}

The coexistence of these heterogeneous vulnerability databases
illustrates the fragmented nature of the vulnerability
information landscape. Automated vulnerability detection tools
must integrate and interpret information originating from
multiple sources that differ in their data models, identifier
schemes, and ecosystem coverage.

\subsection{Software Package Identification}

\paragraph{Terminology.}
We distinguish between \emph{software ecosystems} and
\emph{package registries}. Ecosystems define package naming
schemes, version semantics, and dependency conventions,
whereas registries are concrete services that distribute
packages within these ecosystems. Since the names of major
registries are frequently used as shorthand for their
corresponding ecosystems, we use \texttt{PyPI},
\texttt{npm}, \texttt{NuGet}, and \texttt{Maven} as
ecosystem labels throughout this paper unless a concrete
registry or repository is explicitly referenced.

Identifying affected software components is a central challenge
in automated vulnerability detection. Vulnerability databases
have traditionally described affected software products using
the Common Platform Enumeration (CPE) scheme \cite{cpe}. CPE
provides a standardized naming format for software vendors,
products, and versions and is widely used within the National
Vulnerability Database (NVD) \cite{nvd} to associate
vulnerabilities with software platforms.

While CPE is well suited for identifying software products at
the vendor and product level, it is less suitable for
representing software components distributed through modern
package ecosystems. In ecosystems such as npm, Maven, NuGet, or
PyPI, software components are typically identified using
ecosystem-specific component identifiers combined with explicit
version numbers.

These identifiers represent individual software libraries
rather than complete software products and follow naming
conventions defined by their respective ecosystems. As a
result, mapping between component identifiers and product-level
identifiers such as CPE entries is often ambiguous or
incomplete.

This mismatch between product-oriented identifiers and
package-based software ecosystems introduces challenges for
vulnerability detection systems, which must relate
vulnerability information to the concrete components present
in analyzed applications.

\subsection{Vulnerability Data Model}

Vulnerability information used by automated detection systems
is derived from several interconnected conceptual elements
that describe weaknesses, vulnerability instances, and
affected software components.

Software weaknesses are commonly categorized using the
Common Weakness Enumeration (CWE) taxonomy, which describes
abstract classes of programming errors and security flaws.
Concrete vulnerability instances are identified using
Common Vulnerabilities and Exposures (CVE) identifiers
\cite{cve}. Each CVE entry represents a specific vulnerability
observed in one or more software systems.

Vulnerability databases such as the National Vulnerability
Database (NVD) further associate CVE entries with affected
software products using Common Platform Enumeration (CPE)
identifiers \cite{cpe}. These identifiers provide a
product-oriented representation of affected software.

In contrast, modern advisory systems increasingly describe
vulnerabilities in terms of packages within software
ecosystems. Advisory databases such as GitHub Security
Advisories (GHSA) \cite{ghsa} and the Open Source
Vulnerabilities (OSV) database \cite{osv} link vulnerabilities
to ecosystem identifiers, package names, and affected
version ranges.

Figure~\ref{fig:conceptual-model} illustrates the relationships
between these elements. CWE defines abstract weakness classes,
CVE identifiers represent concrete vulnerability instances,
and CPE provides a product-level abstraction used in
traditional vulnerability databases. Advisory systems
reference vulnerability identifiers and associate them with
ecosystem-specific packages and affected version ranges.

\begin{figure}[t]
\centering
\resizebox{\columnwidth}{!}{%
\begin{tikzpicture}[
  node distance=1.45cm,
  every node/.style={
    draw,
    rectangle,
    rounded corners,
    align=center,
    font=\small,
    text width=3.0cm
  },
  arrow/.style={->, dashed, thick},
  bidir/.style={<->, dashed, thick}
]

\node (ecosystem) {Ecosystem\\(Maven, npm, PyPI)};
\node (component) [below of=ecosystem] {Component / Package};
\node (version) [below of=component] {Version\\(e.g., Semantic Versioning)};

\draw[arrow] (ecosystem) -- (component);
\draw[arrow] (component) -- (version);

\node (cwe) [right=3.0cm of component] {Weakness Class\\(CWE)};
\node (cve) [below of=cwe] {Vulnerability\\(CVE)};

\draw[arrow] (cwe) -- (cve);

\node (cpe) [right=3.2cm of cve] {Product Representation\\(CPE)};
\draw[arrow] (cve) -- (cpe);

\node (advisory) [below=2.0cm of cve] {Advisory\\(GHSA / OSV ID)};

\draw[bidir]
  (cve.south) --
  node[midway, right, font=\scriptsize, draw=none] {references}
  (advisory.north);

\draw[arrow]
  (advisory.west) -- (component.east);

\draw[arrow]
  (advisory.north west) -- (ecosystem.south east);

\draw[arrow]
  (advisory.west) --
  node[midway, left, font=\scriptsize, draw=none] {version range}
  (version.east);

\end{tikzpicture}
}
\caption{Conceptual relationships between weaknesses, vulnerabilities,
advisories, product abstractions, and ecosystem-specific packages.
CWE represents abstract classes of weaknesses, while CVEs identify
concrete vulnerability instances. Traditional vulnerability databases
associate CVEs with software products using CPE identifiers.
Advisory systems reference vulnerabilities and link them to affected
packages within specific ecosystems and version ranges.}
\label{fig:conceptual-model}
\end{figure}

Together, these elements form a heterogeneous vulnerability
information model that vulnerability detection systems must
interpret when determining whether a specific component
version is affected by a vulnerability.

\subsection{Version Semantics and Range Interpretation}

Vulnerability advisories commonly describe affected software
using version ranges rather than enumerating individual vulnerable
versions. Such ranges typically define intervals of affected
versions by specifying lower and upper boundaries, or by
indicating versions in which a vulnerability was fixed.

In modern package ecosystems, version identifiers often follow
ecosystem-specific versioning schemes. Many ecosystems adopt
semantic versioning conventions, where version numbers encode
major, minor, and patch levels of a software release. However,
the interpretation of version ranges may vary depending on
ecosystem rules, advisory formats, and the implementation
details of vulnerability detection systems.

Advisory systems therefore commonly express vulnerability
conditions using range expressions that define intervals of
affected versions. Detection systems must interpret these
expressions in order to determine whether a concrete package
version falls within a vulnerable range.

Table~\ref{tab:version-ranges} illustrates typical forms of
version range specifications used in vulnerability advisories.

\begin{table}[t]
\centering
\caption{Examples of vulnerability version range specifications}
\label{tab:version-ranges}
\begin{tabular}{ll}
\toprule
Expression & Interpretation \\
\midrule
$< 1.4.2$ & All versions prior to 1.4.2 are vulnerable \\
$\geq 1.2.0,\ < 1.3.5$ & Versions between 1.2.0 and 1.3.5 \\
$\leq 2.0.1$ & Versions up to and including 2.0.1 \\
$\geq 3.0.0$ & All versions starting from 3.0.0 are vulnerable \\
\bottomrule
\end{tabular}
\end{table}

In addition to standard release versions, many ecosystems also
use pre-release identifiers as defined by semantic versioning,
such as \texttt{-alpha}, \texttt{-beta}, or \texttt{-rc}. For
example, the version \texttt{1.4.0-alpha.1} represents a
pre-release version preceding the final release \texttt{1.4.0}.
According to semantic versioning rules, pre-release versions are
considered lower than the corresponding final release, e.g.,

\begin{center}
\small\texttt{1.4.0-alpha < 1.4.0-beta < 1.4.0-rc.1 < 1.4.0}.
\end{center}

The interpretation of such versions within vulnerability ranges
may vary across tools. Some detection systems treat pre-release
versions as part of the affected range, while others ignore them
unless they are explicitly specified in an advisory.

As a consequence, two vulnerability detection systems analyzing
the same component version may produce different results even
when relying on the same vulnerability advisory.

\subsection{Implications for Vulnerability Detection}

The heterogeneous structure of vulnerability data sources,
identifier schemes, and version range semantics creates
significant challenges for automated vulnerability detection.

Detection systems must map vulnerability information expressed
in different identifier systems to the software components
present in an analyzed application. In addition, they must
interpret version range specifications in order to determine
whether a specific component version is affected by a
vulnerability.

These steps introduce multiple sources of ambiguity and may
lead to different vulnerability detection results even when
analyzing the same software components. For example, tools may
rely on different vulnerability databases, interpret advisory
information differently, or apply distinct rules when resolving
version ranges.

As a consequence, evaluating and comparing vulnerability
detection systems requires a consistent reference that
describes which component versions are affected by specific
vulnerabilities. Without such a reference, observed
differences between tools cannot be attributed reliably to
their detection logic or to discrepancies in the underlying
vulnerability data sources.

To address this challenge, this paper constructs a curated
ground-truth dataset that explicitly maps vulnerabilities to
concrete package versions across multiple software ecosystems.
The dataset represents a snapshot of the vulnerability
landscape at a specific point in time, since vulnerability
databases continuously evolve as new vulnerabilities are
disclosed and existing records are updated.

Consequently, the dataset must be regenerated for each
evaluation to reflect the current state of the underlying
vulnerability sources. To support this process and enable
reproducible studies, we provide a software tool that
automatically reconstructs the dataset from the available
vulnerability data using the methodology described in this
paper. This approach enables deterministic comparisons of
vulnerability detection results while allowing the dataset to
be updated as vulnerability information evolves.
\section{Ground-Truth Dataset Construction}
\label{sec:dataset}

Evaluating vulnerability detection tools requires a reliable
reference describing which concrete software component
versions are affected by specific vulnerabilities.
To provide such a reference, this work constructs a curated
ground-truth dataset mapping vulnerabilities to explicit
ecosystem–component–version tuples \((e,c,v,u)\), where
\(e\) denotes the ecosystem, \(c\) the component identifier,
\(v\) the concrete component version, and \(u\) the
vulnerability identifier.

The following subsections describe the dataset construction
process, including the extraction of vulnerability records,
the selection of components and versions, the mapping of
vulnerability records to concrete component versions, and the
generation of the final dataset used for the empirical
evaluation.

\subsection{Source Data and Snapshot Extraction}

The ground-truth dataset is derived from vulnerability records
obtained from the Open Source Vulnerabilities (OSV)
database~\cite{osv}.
OSV provides a package-centric vulnerability model that
associates vulnerabilities directly with software ecosystems,
package identifiers, and affected version ranges.

For this study, vulnerability records are retrieved through
the OSV query API for four widely used software dependency
ecosystems: npm, PyPI, Maven, and NuGet.
Each OSV entry contains structured metadata including the
ecosystem identifier, package name, vulnerability identifier
(e.g., CVE or GHSA), and version range specifications
describing affected releases.

Because vulnerability databases evolve continuously as new
vulnerabilities are disclosed and existing records are
updated, the resulting dataset represents a time-bounded
snapshot of the vulnerability landscape at the time of data
collection.
All OSV queries were executed within a fixed data collection
window to ensure that the dataset corresponds to a consistent
snapshot of the OSV database.

OSV entries may reference multiple vulnerability identifiers
(e.g., CVE or GHSA) that describe the same vulnerability.
During dataset construction, identifiers returned by the OSV
API are treated as canonical vulnerability identifiers
\(u\).
If multiple identifiers refer to the same OSV entry, they are
normalized to the identifier returned by the OSV query
response to avoid duplicate vulnerability entries in the
ground-truth dataset.

The retrieved vulnerability records form the basis for
constructing the ground-truth dataset used in this study.

\subsection{Component and Version Selection}

To construct a representative and manageable dataset, the
ground-truth generation process operates on a curated set of
software components for each ecosystem.
For the ecosystems considered in this study (npm, PyPI,
Maven, and NuGet), components were selected from widely used
libraries within the respective ecosystem registries.

The selection focuses on actively maintained packages that
are commonly used as dependencies in real-world software
projects.
The selected components cover several functional domains,
including web frameworks, networking utilities, security
libraries, database clients, and data processing frameworks.

Package identifiers follow the naming conventions of the
respective ecosystems (e.g., \texttt{groupId:artifactId}
coordinates for Maven and canonical package names for npm,
PyPI, and NuGet). Representative examples of analyzed components are shown in
Table~\ref{tab:dataset-components}.

\begin{table}[t]
\centering
\caption{Representative components included in the dataset}
\label{tab:dataset-components}

\footnotesize
\setlength{\tabcolsep}{3pt}
\renewcommand{\arraystretch}{1.1}

\begin{tabular}{
p{1.6cm}
p{6.8cm}
}
\toprule
Ecosystem & Example Components (canonical identifiers) \\
\midrule

Maven &
\texttt{org.apache.logging.log4j},
\texttt{org.springframework:spring-expression} \\

npm &
\texttt{esbuild},
\texttt{vite} \\

NuGet &
\texttt{Microsoft.Data.SqlClient},
\texttt{Microsoft.AspNetCore.Identity} \\

PyPI &
\texttt{requests},
\texttt{keras} \\

\bottomrule
\end{tabular}
\end{table}

For each selected component, a bounded number of recent
stable releases is examined.
Versions are obtained from the respective ecosystem
registries, including the npm registry, the PyPI metadata
API, Maven Central metadata, and the NuGet registry.
Pre-release and development versions are excluded to ensure
that the dataset reflects vulnerabilities affecting stable
production releases.

To control dataset size and maintain balanced ecosystem
coverage, the number of analyzed versions per component is
limited to a fixed upper bound.
For each component, the most recent stable versions are
selected according to the version ordering rules of the
respective ecosystem.

In addition, a configurable time window based on the release
timestamp of each version is applied.
Only component versions whose release dates fall within the
defined interval are considered during dataset construction.

The complete list of analyzed components and versions is
provided in the accompanying dataset construction
repository to enable full reproducibility of the evaluation.
\subsection{Vulnerability--Version Mapping}

OSV vulnerability records typically describe affected software
using version ranges rather than enumerating all vulnerable
versions explicitly.
In this study, the interpretation of version ranges is
delegated to the OSV infrastructure.

For each selected tuple $(e,c,v)$, the OSV query API is invoked
to determine whether the version is affected by known
vulnerabilities.
OSV internally evaluates the affected version ranges
associated with vulnerability records and returns all
vulnerability identifiers applicable to the queried version.

Each returned vulnerability identifier $u$ results in a
ground-truth entry $(e,c,v,u)$.
This representation links vulnerabilities directly to
explicit component versions rather than abstract version
ranges.

The resulting dataset therefore preserves the one-to-many
relationship between component versions and vulnerability
records.
A single component version may correspond to multiple
ground-truth entries when several distinct vulnerabilities
affect the same release.

Overall, the dataset captures vulnerability information across
multiple dependency ecosystems.
Each entry represents a mapping between a vulnerability
identifier, a component identifier, and a concrete component
version affected by that vulnerability.
\subsection{Tool Support and Dataset Regeneration}

Since vulnerability databases evolve continuously as new
vulnerabilities are disclosed and existing records are updated,
any dataset derived from such sources represents only a
snapshot of the vulnerability landscape at a specific point in
time.

To support repeated evaluations under evolving vulnerability
data, we developed a software tool that automates the dataset
construction workflow described above. The tool extracts
vulnerability records from the OSV database, retrieves
component versions from ecosystem registries, and generates
the normalized representation linking vulnerabilities to
concrete component versions.

By automating the dataset construction process, the ground
truth dataset can be regenerated for different snapshots of
the underlying vulnerability data while preserving the
methodology described in this work.

\section{Evaluation Methodology}
\label{sec:evaluation}

This section describes the methodology used to evaluate
vulnerability detection tools using the ground-truth dataset
constructed in Section~\ref{sec:dataset}. The evaluation
operates on explicit ecosystem--component--version tuples,
compares normalized tool-reported vulnerabilities with the
ground-truth using deterministic matching semantics, and
controls for temporal drift through repeated executions on
stable ground-truth snapshots.

\subsection{Evaluation Scope and Input Contract}
\label{subsec:eval-scope}

All evaluated tools operate on a fixed set of concrete
software dependencies represented as tuples
\[
(e,c,v),
\]
where \(e\) denotes the software ecosystem,
\(c\) the component identifier, and
\(v\) the exact component version.

Each tuple therefore represents a single dependency instance
that is evaluated independently. By fixing the evaluation
input to explicit ecosystem--component--version tuples, the
methodology isolates vulnerability resolution from other
aspects of dependency analysis, such as dependency discovery,
build configuration, or project-specific scanning heuristics.

Only vulnerabilities affecting these concrete dependency
instances are considered relevant for the evaluation.
Vulnerabilities reported for other components or versions are
treated as over-approximations and are therefore classified
as ground-truth-relative false positives (\(FP_{GT}\)).

\subsection{Ground Truth Matching Semantics}
\label{subsec:ground-truth-matching}

To evaluate vulnerability detection tools, normalized tool
results are compared against the OSV-derived ground-truth
dataset. Each ground-truth entry is represented as
\[
(e,c,v,u),
\]
where \(e\) denotes the software ecosystem,
\(c\) the component identifier,
\(v\) the component version, and
\(u\) the vulnerability identifier.

Let \(GT\) denote the set of ground-truth vulnerabilities
affecting the evaluated component-version tuples and
\(R_t\) the set of vulnerabilities reported by tool \(t\).

\begin{figure}[t]
\centering
\begin{tikzpicture}[scale=0.8, every node/.style={transform shape}]

\tikzstyle{box}=[
draw,
rounded corners,
align=center,
text width=3.2cm,
minimum height=0.9cm
]

\tikzstyle{arrow}=[->, thick]

\node[box] (input) at (0,0) {Input Set\\$I=\{(e,c,v)\}$};
\node[box] (tool) at (0,-1.8) {Tool $t$\\Vulnerability\\Detection};
\node[box] (results) at (0,-3.6) {Tool Results\\$R_t$};

\node[box] (gt) at (4.5,-1.8) {Ground Truth\\Dataset\\$GT=\{(e,c,v,u)\}$};
\node[box] (match) at (4.5,-3.6) {Matching};
\node[box] (metrics) at (4.5,-5.4) {Evaluation\\Metrics};

\draw[->, thick] (input) -- (tool);
\draw[->, thick] (tool) -- (results);
\draw[->, thick] (results) -- (match);
\draw[->, thick] (gt) -- (match);
\draw[->, thick] (match) -- (metrics);

\end{tikzpicture}
\caption{Overview of the evaluation methodology. The input set $I$ defines the evaluated dependency instances. A vulnerability detection tool $t$ produces results $R_t$, which are compared with the ground-truth dataset $GT$. Matching yields the sets of true positives, ground-truth-relative false positives, and false negatives, from which evaluation metrics are derived.}
\label{fig:evaluation-methodology}
\end{figure}
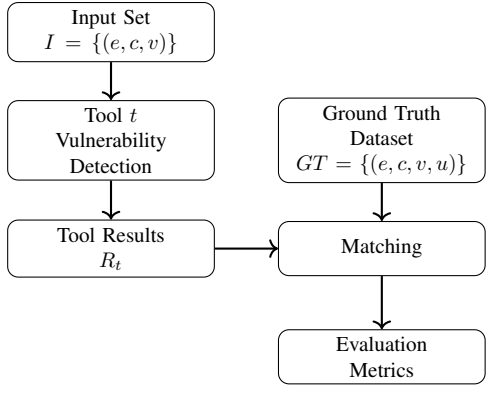

Figure~\ref{fig:evaluation-methodology} summarizes the
overall evaluation methodology used in this study.
Ground-truth entries derived from OSV are compared with
normalized vulnerability reports produced by each tool.

\begin{figure}[htbp]
\centering
\begin{tikzpicture}[scale=0.9]

\draw (0,0) circle (1.2);
\draw (1.2,0) circle (1.2);

\begin{scope}
\clip (0,0) circle (1.2);
\fill[gray!20] (1.2,0) circle (1.2);
\end{scope}

\node at (-0.45,0) {$FN$};
\node at (0.6,0) {$TP$};
\node at (1.9,0) {$FP_{GT}$};

\node at (-0.7,1.5) {$GT$};
\node at (2.0,1.5) {$R_t$};

\end{tikzpicture}
\caption{Relationship between the ground-truth set $GT$ and the result set $R_t$ reported by tool $t$.}
\label{fig:matching-sets}
\end{figure}
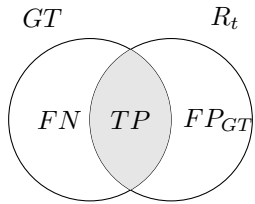

The relationship between these sets is illustrated in
Figure~\ref{fig:matching-sets}.

True positives correspond to vulnerabilities correctly
reported by the tool:
\begin{equation}
TP = GT \cap R_t
\end{equation}

Ground-truth-relative false positives correspond to
vulnerabilities reported by the tool but not present in the
ground-truth:
\begin{equation}
FP_{GT} = R_t \setminus GT
\end{equation}

False negatives correspond to vulnerabilities present in the
ground-truth but not detected by the tool:
\begin{equation}
FN = GT \setminus R_t
\end{equation}

A reported ground-truth-relative false positive does not
necessarily imply that the reported vulnerability is
incorrect. It may also correspond to a legitimate
vulnerability that was not yet included in OSV at the time
the ground-truth snapshot was constructed, or to a finding
that is not represented in the reference dataset due to
reference incompleteness or update lag.

Based on these sets, the following evaluation metrics are
used.

Recall measures the fraction of ground-truth vulnerabilities
that are correctly detected by the tool:
\begin{equation}
Recall = \frac{|TP|}{|TP| + |FN|}
\end{equation}

Overlap measures the proportion of reported vulnerabilities
that coincide with the ground-truth:
\begin{equation}
Overlap = \frac{|TP|}{|TP| + |FP_{GT}|}
\end{equation}

While recall quantifies the coverage of a tool with respect
to the ground-truth, overlap captures the degree to which
tool findings agree with the current reference dataset and is
therefore interpreted relative to the ground-truth snapshot.

\subsection{Evaluation Procedure}
\label{subsec:evaluation-procedure}

The evaluation procedure applies each vulnerability detection
tool to the same set of component-version tuples derived from
the ground-truth dataset. Each tuple represents a concrete
dependency instance of the form \((e,c,v)\), where \(e\)
denotes the ecosystem, \(c\) the component identifier, and
\(v\) the exact component version.

For each evaluated component version, the corresponding tool
is executed or queried to obtain the set of vulnerabilities
that the tool considers applicable to the analyzed dependency.
The resulting vulnerability reports form the tool result set
\(R_t\).

Since vulnerability detection tools expose their results
through heterogeneous interfaces and report formats, tool
outputs are processed through tool-specific adapters that
retrieve vulnerability findings from the respective tool APIs
or report exports and transform them into a unified internal
representation. Each normalized finding records the ecosystem,
component identifier, component version, and vulnerability
identifier associated with the reported issue.

This adapter-based normalization step ensures that the
subsequent evaluation operates on a consistent representation
independent of the original tool output format and allows
different tools to be compared using a common evaluation
pipeline.

The normalized tool findings are then compared against the
ground-truth dataset according to the matching semantics
defined in the previous subsection. Based on this comparison,
each reported finding is classified as a true positive (\(TP\)),
ground-truth-relative false positive (\(FP_{GT}\)), or false
negative (\(FN\)). True positives may occur either as exact
matches, where the tool reports the same component version
and vulnerability identifier as the ground-truth, or as
range-based matches when a tool reports a vulnerability
associated with a version interval that covers the evaluated
component version.

The resulting sets of true positives, ground-truth-relative
false positives, and false negatives form the basis for the
quantitative evaluation metrics introduced earlier in this
section.

\subsection{Temporal Consistency Control}
\label{subsec:temporal-consistency}

Since vulnerability databases evolve continuously, any
dataset derived from them reflects only a snapshot of the
vulnerability landscape at a specific point in time. New
vulnerabilities may be added, existing records may be
revised, and metadata may change while an experiment is
running. Without temporal control, tool comparisons may
therefore be distorted by changes in the underlying data
rather than by actual differences in tool behavior.

To mitigate this effect, the evaluation follows a temporally
controlled workflow that accepts only runs executed on a
stable ground-truth snapshot. For each attempt, a
ground-truth snapshot is constructed before tool execution
and reconstructed again after all tool runs have finished.
The attempt is accepted only if both snapshots are identical
and all tool executions complete successfully. In practice,
snapshot equality is verified via a hash of the generated
ground-truth representation.

The procedure is illustrated in
Algorithm~\ref{alg:temporal-consistency}. First, an initial
ground-truth snapshot \(GT_0\) is created. Next, all
evaluated tools are executed twice on the same fixed
snapshot. For each repeat, result hashes are recorded in
order to document whether repeated executions produced
identical tool outputs. These repeat hashes are recorded and
inspected manually before results are included in the final
analysis in order to verify that repeated executions
remained consistent. Afterwards, the ground-truth is rebuilt
as \(GT_1\). If the hash of \(GT_0\) differs from the hash
of \(GT_1\), the attempt is discarded and repeated.

If the attempt is accepted, repeat-level metrics are
aggregated for reporting. In addition, binary detection
vectors from both repeats are concatenated for the
statistical comparison. This separates automated temporal
consistency control from the additional manual inspection of
repeat stability.

\begin{algorithm}[t]
\caption{Temporal Consistency Controlled Evaluation}
\label{alg:temporal-consistency}
\begin{algorithmic}[1]
\Require tool set $T$, maximum attempts $A_{\max}$, repeats $R=2$

\For{$a \gets 1$ to $A_{\max}$}
    \State $(GT_0, SBOM_0) \gets \mathrm{BuildGT}()$
    \State $h_0 \gets \mathrm{HashGT}(GT_0)$
    \State run all tools for $R$ repeats on $(GT_0, SBOM_0)$
    \If{any execution fails}
        \State \textbf{continue}
    \EndIf
    \State record repeat hashes for later consistency inspection
    \State $(GT_1, SBOM_1) \gets \mathrm{BuildGT}()$
    \State $h_1 \gets \mathrm{HashGT}(GT_1)$
    \If{$h_0 \neq h_1$}
        \State \textbf{continue}
    \EndIf
    \State aggregate repeat-level metrics
    \State concatenate binary detection vectors across repeats
    \State compute overall and pairwise significance tests
    \State \textbf{accept run and break}
\EndFor

\If{no attempt was accepted}
    \State \textbf{abort}
\EndIf
\end{algorithmic}
\end{algorithm}

Together, these steps provide a systematic and temporally
controlled basis for comparing vulnerability detection
behavior across tools and ecosystems.

\subsection{Artifact Availability}
\label{sec:artifact-availability}

To support transparency and reproducibility, we provide both
the evaluation tool and the generated ground-truth dataset as
publicly archived research artifacts. The evaluation tool is
released as open-source software and archived in a versioned
repository snapshot. The ground-truth dataset is published as
a separate versioned dataset artifact, including the dataset
files, generation parameters, and summary statistics used in
this study.

For long-term reference, we cite the concept-level persistent
identifiers of both artifacts. For exact reproducibility of
the results reported in this paper, we additionally reference
the specific archived versions that were used for the
experiments. The software artifact contains the source code,
documentation, and configuration required to reproduce the
evaluation workflow. The dataset artifact contains the
ground-truth files, schema information, and accompanying
descriptive statistics.

\medskip
\noindent\textbf{Software artifact for the tool:}
\emph{Evaluation Tool for SCA Benchmarking}, version DOI:
\url{https://doi.org/10.5281/zenodo.19697250}.

\noindent\textbf{Dataset artifact for the tool:}
\emph{Ground-Truth Dataset for SCA Tool Evaluation}, version DOI:
\url{https://doi.org/10.5281/zenodo.19696658}.

\section{Vulnerability Detection Tools}
\label{sec:tools}

Modern software composition analysis (SCA) tools rely on a
variety of vulnerability databases and differ in their
approaches to dependency analysis, vulnerability aggregation,
and version interpretation. As a result, the vulnerabilities
reported for a given software component may vary significantly
across tools even when analyzing the same dependency version.

\begin{table*}[!t]
\centering
\caption{Overview of representative vulnerability detection and advisory tools}
\label{tab:vuln-tools-overview}
\small
\setlength{\tabcolsep}{2.5pt}
\renewcommand{\arraystretch}{1.15}

\begin{tabular}{@{}
>{\raggedright\arraybackslash}p{0.17\textwidth}
>{\raggedright\arraybackslash}p{0.12\textwidth}
>{\raggedright\arraybackslash}p{0.18\textwidth}
>{\raggedright\arraybackslash}p{0.37\textwidth}
>{\centering\arraybackslash}p{0.07\textwidth}
@{}}
\toprule
Tool / Service & Type & Primary Data Sources & Description & Ref. \\
\midrule

Black Duck &
Commercial SCA platform &
Black Duck KB, CVE, NVD &
Commercial software composition analysis platform maintained by
Synopsys providing vulnerability detection and governance. &
\cite{blackduck} \\

Dependabot &
Integrated monitoring service &
GitHub Advisory DB &
Automated dependency monitoring service integrated into GitHub
that generates alerts when vulnerable dependencies are detected. &
\cite{dependabot} \\

Dependency-Track &
Open-source SCA platform &
NVD, OSS Index, GHSA &
SBOM-driven software composition analysis platform aggregating
vulnerability information from multiple advisory sources. &
\cite{dependencytrack} \\

FOSSA &
Commercial SCA platform &
Multiple advisory databases &
Commercial platform providing vulnerability detection and
open-source license compliance analysis. &
\cite{fossa} \\

GitHub Advisory Database &
Advisory service &
GHSA, CVE &
Package-level vulnerability advisory database maintained by
GitHub. &
\cite{ghsa} \\

Grype &
Open-source scanner &
NVD, GitHub Advisories, vendor advisories &
Open-source vulnerability scanner focusing on container images
and application dependencies. &
\cite{grype} \\

Mend (WhiteSource) &
Commercial SCA platform &
Mend vulnerability DB, CVE &
Commercial dependency security platform providing vulnerability
detection and open-source governance capabilities. &
\cite{mend} \\

OSS Index &
Advisory service &
OSS Index DB &
Curated vulnerability advisory service mapping vulnerabilities
to package coordinates across multiple ecosystems. &
\cite{ossindex} \\

OSV-Scanner &
Open-source scanner &
OSV &
Lightweight scanner that queries the OSV database directly to
identify vulnerabilities affecting dependency versions. &
\cite{osvscanner} \\

Snyk &
Commercial SCA platform &
Snyk DB, CVE, ecosystem advisories &
Commercial dependency vulnerability scanner combining public
vulnerability databases with curated vulnerability intelligence. &
\cite{snyk} \\

Trivy &
Open-source scanner &
NVD, OSV, vendor advisories &
Security scanner capable of detecting vulnerabilities in
containers, infrastructure configurations, and software
dependencies. &
\cite{trivy} \\

\bottomrule
\end{tabular}
\end{table*}

\subsection{Vulnerability Detection Ecosystem}

\begin{figure}[t]
\centering
\resizebox{0.65\columnwidth}{!}{
\begin{tikzpicture}[
node distance=1.1cm,
box/.style={
draw,
rounded corners,
align=center,
font=\tiny,
minimum width=3.4cm,
minimum height=0.6cm
},
arrow/.style={->, thick}
]

\node[box] (db)
{Vulnerability Databases\\
NVD, OSV, CVE};

\node[box, below of=db] (advisory)
{Advisory Services\\
GitHub Advisories, OSS Index};

\node[box, below of=advisory] (tools)
{Scanning / SCA Tools\\
Snyk, Trivy, Dependency-Track, Grype};

\draw[arrow] (db.south) -- (advisory.north);
\draw[arrow] (advisory.south) -- (tools.north);

\end{tikzpicture}
}
\caption{Simplified architecture of vulnerability detection
ecosystems. Vulnerability databases provide vulnerability
records, advisory services aggregate and map these records to
software ecosystems, and scanning tools consume this
information to detect vulnerabilities in software dependencies.}
\label{fig:vuln-ecosystem}
\end{figure}
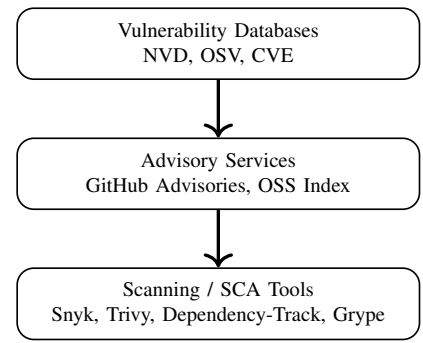

Automated vulnerability detection operates within a broader
ecosystem consisting of vulnerability databases, advisory
aggregation services, and vulnerability scanning tools.
Each layer provides different types of information required
to identify vulnerabilities affecting software dependencies.
Figure~\ref{fig:vuln-ecosystem} summarizes this layered
vulnerability detection ecosystem.

At the foundation of this ecosystem are vulnerability
databases such as CVE, NVD, and OSV, which provide identifiers
and structured metadata describing publicly disclosed
vulnerabilities.

On top of these databases, advisory aggregation services
collect and normalize vulnerability information for specific
software ecosystems. Services such as GitHub Advisories and
OSS Index map vulnerability identifiers to ecosystem-specific
package identifiers and affected version ranges.

Finally, vulnerability scanning tools and SCA platforms
consume this information in order to detect vulnerabilities
affecting dependencies used in software projects.

\subsection{Vulnerability Data Sources and Access Mechanisms}

\begin{figure}[t]
\centering
\resizebox{0.9\columnwidth}{!}{%
\begin{tikzpicture}[
arrow/.style={->, thick},
topbox/.style={
draw,
rounded corners,
align=center,
font=\small,
text width=2.4cm,
minimum height=0.9cm
},
midbox/.style={
draw,
rounded corners,
align=center,
font=\small,
text width=3.2cm,
minimum height=1.0cm
},
botbox/.style={
draw,
rounded corners,
align=center,
font=\small,
text width=2.1cm,
minimum height=0.9cm
}
]

\node[topbox] (cve) at (-2.3,0) {CVE\\(MITRE)};
\node[topbox] (nvd) at ( 2.3,0) {NVD\\(NIST)};

\node[midbox] (osv) at (0,-2.2) {OSV\\(Aggregation \&\\Normalization)};

\node[botbox] (PyPI)  at (-4.5,-4.8) {PyPI\\Advisories};
\node[botbox] (npm)   at (-1.5,-4.8) {npm\\Advisories};
\node[botbox] (Maven) at ( 1.5,-4.8) {Maven\\Advisories};
\node[botbox] (NuGet) at ( 4.5,-4.8) {NuGet\\Advisories};

\draw[arrow] (cve.south) -- ([xshift=-0.9cm]osv.north);
\draw[arrow] (nvd.south) -- ([xshift= 0.9cm]osv.north);

\draw[arrow] (PyPI.north)  -- ([xshift=-1.2cm]osv.south);
\draw[arrow] (npm.north)   -- ([xshift=-0.4cm]osv.south);
\draw[arrow] (Maven.north) -- ([xshift= 0.4cm]osv.south);
\draw[arrow] (NuGet.north) -- ([xshift= 1.2cm]osv.south);

\end{tikzpicture}
}
\caption{Simplified conceptual data flow between vulnerability
identifier catalogs, normalized advisory datasets, and
ecosystem-specific repositories.}
\label{fig:vuln-data-flow}
\end{figure}
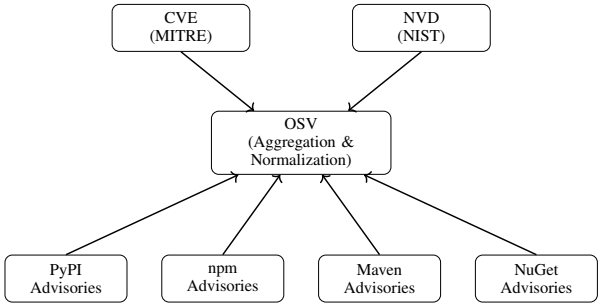

Figure~\ref{fig:vuln-data-flow} illustrates a simplified
conceptual view of how vulnerability identifier catalogs,
normalized advisory datasets, and ecosystem-specific package
advisories relate to one another.

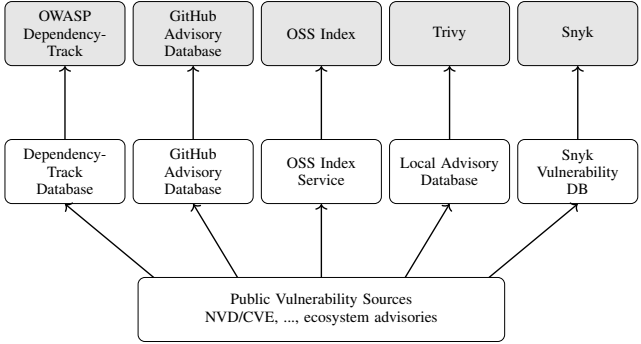
\begin{figure}[t]
\centering
\resizebox{0.96\columnwidth}{!}{%
\begin{tikzpicture}[
every node/.style={align=center, font=\small},
source/.style={
draw,
rounded corners,
text width=7.2cm,
minimum height=1.3cm
},
service/.style={
draw,
rounded corners,
text width=2.2cm,
minimum height=1.3cm
},
tool/.style={
draw,
rounded corners,
fill=black!10,
text width=2.2cm,
minimum height=1.3cm
},
arrow/.style={->, thick}
]

\node[source] (public) at (0,0)
{Public Vulnerability Sources\\NVD/CVE, ..., ecosystem advisories};

\node[service] (dtrackdb) at (-5.2,2.8) {Dependency-Track\\Database};
\node[service] (ghsadb)   at (-2.6,2.8) {GitHub Advisory\\Database};
\node[service] (ossdb)    at ( 0.0,2.8) {OSS Index\\Service};
\node[service] (trivydb)  at ( 2.6,2.8) {Local Advisory\\Database};
\node[service] (snykdb)   at ( 5.2,2.8) {Snyk\\Vulnerability DB};

\node[tool] (dtrack) at (-5.2,5.6) {OWASP\\Dependency-Track};
\node[tool] (ghadv)  at (-2.6,5.6) {GitHub Advisory\\Database};
\node[tool] (ossidx) at ( 0.0,5.6) {OSS Index};
\node[tool] (trivy)  at ( 2.6,5.6) {Trivy};
\node[tool] (snyk)   at ( 5.2,5.6) {Snyk};

\draw[arrow] (-3.4,0.65) -- (dtrackdb.south);
\draw[arrow] (-1.7,0.65) -- (ghsadb.south);
\draw[arrow] ( 0.0,0.65) -- (ossdb.south);
\draw[arrow] ( 1.7,0.65) -- (trivydb.south);
\draw[arrow] ( 3.4,0.65) -- (snykdb.south);

\draw[arrow] (dtrackdb.north) -- (dtrack.south);
\draw[arrow] (ghsadb.north)   -- (ghadv.south);
\draw[arrow] (ossdb.north)    -- (ossidx.south);
\draw[arrow] (trivydb.north)  -- (trivy.south);
\draw[arrow] (snykdb.north)   -- (snyk.south);

\end{tikzpicture}
}
\caption{Conceptual view of how the evaluated tools access
and organize vulnerability information. The figure abstracts
from tool-specific ingestion pipelines and highlights only
the high-level relationship between public vulnerability
sources, intermediate services or internal databases, and
tool-level consumption.}
\label{fig:tools-databases}
\end{figure}

While Figure~\ref{fig:vuln-data-flow} focuses on the broader
vulnerability data landscape, Figure~\ref{fig:tools-databases}
illustrates at a high level how the evaluated tools access
and organize vulnerability information.

The figure intentionally abstracts from tool-specific
ingestion pipelines. In practice, individual tools and
services may aggregate, mirror, or transform vulnerability
information from public sources such as NVD/CVE, GHSA, OSV,
and ecosystem-specific advisories.

\subsection{Tool Integration and Result Normalization}

To enable a systematic comparison across tools, each evaluated
tool is integrated through a tool-specific adapter that
retrieves vulnerability findings from tool APIs or report
exports and transforms them into a unified internal
representation.

Table~\ref{tab:tool-access} summarizes the access mechanisms,
required inputs, and invocation patterns of the evaluated
tools.

\begin{table*}[t]
\centering
\caption{Tool access mechanisms and invocation patterns}
\label{tab:tool-access}
\small
\setlength{\tabcolsep}{4pt}
\renewcommand{\arraystretch}{1.1}

\begin{tabular}{
p{3.8cm}
p{2.8cm}
p{3.0cm}
p{6.6cm}
}
\toprule
Tool & Access Mode & Required Input & Invocation Pattern \\
\midrule

OWASP Dependency-Track &
REST API &
CycloneDX SBOM &
SBOM upload followed by server-side vulnerability analysis. \\

Snyk &
REST API / CLI &
CycloneDX SBOM or package coordinates &
SBOM or package metadata submitted to the Snyk backend. \\

Sonatype OSS Index &
REST API &
Package coordinates (purl, version) &
Batch submission of component coordinates via REST API. \\

GitHub Advisory Database &
GraphQL API &
Package name and version &
Per-component advisory queries against GitHub endpoints. \\

Trivy &
CLI &
Local project or SBOM &
Local dependency scanning using aggregated advisory databases. \\

\bottomrule
\end{tabular}
\end{table*}

The normalized tool findings are subsequently compared
against the ground-truth dataset according to the matching
semantics defined in
Section~\ref{sec:evaluation}, in particular
Subsection~\ref{subsec:ground-truth-matching}.


\subsection{Tool Landscape and Evaluated Systems}

Building on the vulnerability data ecosystem outlined above,
Table~\ref{tab:vuln-tools-overview} summarizes representative
vulnerability detection tools and advisory services widely
used in practice.

From this broader ecosystem, only a subset of systems is
included in the empirical evaluation conducted in this study.
The evaluated tools are \textbf{Snyk} as a representative
commercial SCA platform, \textbf{OWASP Dependency-Track} as a
widely used open-source SCA platform, \textbf{Trivy} as a
lightweight open-source vulnerability scanner, and two
advisory lookup services, namely \textbf{Sonatype OSS Index}
and the \textbf{GitHub Advisory Database}.

These tools were selected based on three criteria.
First, they represent different architectural approaches to
vulnerability detection. Commercial SCA platforms such as
Snyk combine multiple vulnerability sources with proprietary
curation and enrichment processes. Open-source SCA platforms
such as Dependency-Track aggregate vulnerability information
from several public advisory databases and integrate external
analyzers. Lightweight scanners such as Trivy perform
dependency analysis directly within local environments using
aggregated vulnerability data. Advisory lookup services such
as OSS Index and the GitHub Advisory Database provide direct
access to curated vulnerability information without performing
full project analysis.

Second, the selected systems rely on different vulnerability
data sources and aggregation strategies. Some tools combine
multiple public vulnerability databases such as NVD, GHSA,
and OSV, while others rely primarily on curated proprietary
datasets. Evaluating tools with heterogeneous data provenance
makes it possible to study how differences in vulnerability
data aggregation and normalization affect detection results.

Third, all selected systems provide programmatic access
through command-line interfaces or public APIs. This property
is essential for enabling automated and reproducible
evaluations across a large number of component versions.

By selecting tools from these categories, the evaluation
captures representative differences in vulnerability data
aggregation, version interpretation, and vulnerability
resolution strategies across the current vulnerability
detection ecosystem.

\section{Experiments and Results}
\label{sec:experiments}

This section evaluates vulnerability resolution across tools
using the OSV-derived ground-truth dataset introduced in
Section~\ref{sec:dataset}. All experiments are conducted on
the same fixed set of explicit ecosystem--component--version
inputs and use the identical matching semantics defined in
Subsection~\ref{subsec:ground-truth-matching}. This ensures
that observed differences reflect vulnerability resolution
behavior rather than differences in dependency discovery or
project configuration.

The evaluation covers the four ecosystems Maven, npm, PyPI,
and NuGet, and includes OWASP Dependency-Track, Snyk,
Sonatype OSS Index, the GitHub Advisory Database, and Trivy.
The goal is to measure how closely these tools reproduce the
OSV-derived reference set of vulnerable dependency instances
when given identical package and version inputs. OSV itself
serves as the reference source and, by construction, achieves
perfect agreement with the OSV-derived ground-truth.

The experiments were conducted on macOS Sequoia 15.7.4 on an
iMac (2019), equipped with a 3.6\,GHz
8-core Intel Core i9 processor, 32\,GB DDR4 memory, and a
Radeon Pro 580X GPU with 8\,GB VRAM. The accepted temporally
controlled run was carried out on 28 March 2026 between
16:27:09 and 20:51:14. It consisted of one ground-truth
reconstruction before execution, two repeated tool passes on
the same fixed input set, and one ground-truth reconstruction
after execution. Both repeats produced identical tool outputs,
and the before/after ground-truth hashes matched, indicating
that no relevant snapshot drift occurred during the execution
window.


\subsection{Characteristics of the Generated Ground-Truth Dataset}
\label{subsec:dataset-characteristics}

To contextualize the evaluation results, this subsection
summarizes both the overall composition and the structural
characteristics of the generated ground-truth dataset. Across
the four ecosystems \texttt{Maven}, \texttt{npm},
\texttt{NuGet}, and \texttt{PyPI}, the dataset contains
430 unique components, 1000 OSV vulnerability entries,
924 CVE-backed findings, and 189 distinct CVE identifiers.
The generated ground-truth dataset and the evaluation tool are
publicly released as versioned research artifacts; details are
provided in Section~\ref{sec:artifact-availability}.

Table~\ref{tab:dataset-detailed-main} reports the detailed
per-ecosystem composition and structural ratio measures,
whereas Table~\ref{tab:dataset-frequency-main} reports the
component and component-version frequency statistics. Although
the dataset is perfectly balanced with respect to the number
of OSV vulnerability entries per ecosystem (250 each), it is
not balanced with respect to component diversity, CVE
diversity, or concentration patterns. This distinction is
important because equal numbers of vulnerability entries do
not necessarily imply equal structural difficulty for SCA
tools.

In Table~\ref{tab:dataset-detailed-main}, \textit{Comp.}
denotes the number of unique components, \textit{OSV} the
number of OSV vulnerability entries, \textit{CVE-F.} the
number of CVE-backed findings, and \textit{CVEs} the number of
distinct CVE identifiers. In addition, \textit{V-Share}
denotes the share of all OSV vulnerability entries and
\textit{C-Share} the share of all unique components. Two ratio
measures further highlight structural differences between
ecosystems. \textit{Comp./OSV} is the ratio of unique
components to OSV vulnerability entries. Higher values
indicate that the same number of OSV entries is distributed
across a broader set of components, whereas lower values
indicate stronger concentration on relatively few components.
Accordingly, \texttt{NuGet} has the highest value
(\(0.76\)), while \texttt{npm} (\(0.26\)) and \texttt{PyPI}
(\(0.30\)) are much more concentrated. \textit{CVE-F./OSV} is
the ratio of CVE-backed findings to OSV entries and shows how
many OSV entries could be mapped to explicit CVE-based
ground-truth findings. \texttt{NuGet} reaches \(1.00\),
indicating that all OSV entries in this ecosystem are
CVE-backed, whereas \texttt{PyPI} has the lowest value
(\(0.81\)).

Table~\ref{tab:dataset-frequency-main} reports the frequency
distribution of components and component-version pairs. A
\textit{component frequency} measures how often a component
appears in the dataset, regardless of version. A
\textit{component-version frequency} measures how often the
same concrete component-version pair occurs. The statistics
\textit{Max-C}, \textit{Avg-C}, \textit{Min-C}, and
\textit{Med-C} therefore denote the maximum, average,
minimum, and median component frequency, respectively.
Analogously, \textit{Max-CV}, \textit{Avg-CV},
\textit{Min-CV}, and \textit{Med-CV} describe the same
statistics for component-version frequencies. The strongest
concentration can be observed in \texttt{npm}. Its
\textit{Max-C} value of 200 means that the most frequently
occurring component is associated with 200 OSV entries, and
its average component frequency (\textit{Avg-C} \(= 31.25\))
is also clearly higher than in the other ecosystems. This
indicates that a comparatively small number of components
accounts for a large portion of the vulnerability entries. By
contrast, \texttt{NuGet} is structurally broader: it contains
the largest number of unique components, but has a much lower
average component frequency (\(9.26\)) and a maximum
component-version frequency of only 3. \texttt{Maven} also
shows notable concentration, whereas \texttt{PyPI} lies
between these extremes.

Beyond per-ecosystem size, the dataset therefore exhibits
substantial structural heterogeneity. In particular,
\texttt{npm} is characterized by strong concentration on a
relatively small set of highly recurring components, whereas
\texttt{NuGet} distributes the same number of OSV entries
across a much larger set of components. \texttt{PyPI}, in
turn, combines comparatively high CVE diversity with lower
advisory-to-CVE consolidation. These differences indicate
that the evaluation does not compare tools on four
structurally equivalent subdatasets, but on four ecosystems
that expose distinct matching and identification challenges.

A second important aspect is the relation between OSV entries
and CVE-backed findings. The dataset contains substantially
fewer distinct CVE identifiers than OSV vulnerability
entries, which implies that multiple OSV records may map to
the same underlying CVE. Consequently, equal dataset size in
terms of OSV entries should not be interpreted as equal
identifier diversity. This distinction is relevant for the
interpretation of tool performance, because tools that
primarily rely on CVE-centric matching may face different
conditions from tools that operate more directly on advisory-
or package-level information.

A temporally shifted comparison with a later regenerated
ground-truth snapshot further indicates that the dataset is
structurally stable at a coarse level, but not fully
time-invariant in its concrete vulnerability assignments. In
the repeated run, the overall dataset size and the
per-ecosystem OSV distribution remained unchanged, whereas
small changes occurred in CVE-backed findings and individual
ground-truth entries, with visible effects on several tool
results. The key finding is therefore that reproducibility is
high at the structural level, but evaluation outcomes still
depend to some extent on the temporal state of the underlying
advisory data. A detailed comparison of both snapshots and the
resulting tool-level differences is provided in
Appendix~\ref{app:temporal-gt-eval-comparison}.

Finally, the ground-truth construction is shaped by explicit
generation parameters, including ecosystem-specific caps on
the number of versions considered per package and a common
release-date window. These design choices improve
reproducibility, but they also influence the structural
composition of the resulting dataset. For this reason, the
reported results should be interpreted relative to the present
ground-truth configuration rather than as ecosystem-
independent performance estimates.

\begin{table*}[t]
\centering
\caption{Detailed per-ecosystem statistics of the generated ground-truth dataset.}
\label{tab:dataset-detailed-main}
\small
\setlength{\tabcolsep}{3.2pt}
\renewcommand{\arraystretch}{1.08}
\begin{tabular*}{\textwidth}{@{\extracolsep{\fill}}lrrrrrrrr@{}}
\toprule
Eco & Comp. & OSV & CVE-F. & CVEs & Comp./OSV & CVE-F./OSV & V-Share & C-Share \\
\midrule
Maven & 99  & 250  & 240 & 42  & 0.40 & 0.96 & 25.0\%  & 23.0\% \\
npm   & 66  & 250  & 231 & 19  & 0.26 & 0.92 & 25.0\%  & 15.4\% \\
NuGet & 189 & 250  & 250 & 36  & 0.76 & 1.00 & 25.0\%  & 44.0\% \\
PyPI  & 76  & 250  & 203 & 92  & 0.30 & 0.81 & 25.0\%  & 17.7\% \\
\midrule
\textbf{TOTAL} & 430 & 1000 & 924 & 189 & 0.43 & 0.92 & 100.0\% & 100.0\% \\
\bottomrule
\end{tabular*}

\smallskip
\raggedright
\footnotesize
Abbreviations: Eco = ecosystem, Comp. = unique components,
OSV = OSV vulnerability entries, CVE-F. = CVE-backed findings,
CVEs = distinct CVE identifiers, Comp./OSV = ratio of unique
components to OSV vulnerability entries, CVE-F./OSV = ratio of
CVE-backed findings to OSV vulnerability entries, V-Share =
share of OSV vulnerability entries, C-Share = share of unique
components. \textit{Note:} The CVE count in the TOTAL row is
reported as the sum of the per-ecosystem CVE counts.
\end{table*}

\begin{table*}[t]
\centering
\caption{Component and component-version frequency statistics of the generated ground-truth dataset.}
\label{tab:dataset-frequency-main}
\small
\setlength{\tabcolsep}{3.2pt}
\renewcommand{\arraystretch}{1.08}
\begin{tabular*}{\textwidth}{@{\extracolsep{\fill}}lrrrrrrrr@{}}
\toprule
Eco & Max-C & Avg-C & Min-C & Med-C & Max-CV & Avg-CV & Min-CV & Med-CV \\
\midrule
Maven & 42  & 16.67 & 1 & 10.00 & 10 & 2.53 & 1 & 2.00 \\
npm   & 200 & 31.25 & 2 & 6.50  & 10 & 3.79 & 1 & 1.00 \\
NuGet & 33  & 9.26  & 1 & 8.00  & 3  & 1.32 & 1 & 1.00 \\
PyPI  & 30  & 7.58  & 1 & 4.00  & 10 & 3.29 & 1 & 2.00 \\
\midrule
\textbf{TOTAL} & 200 & 12.05 & 1 & 6.00 & 10 & 2.33 & 1 & 1.00 \\
\bottomrule
\end{tabular*}

\smallskip
\raggedright
\footnotesize
Abbreviations: Eco = ecosystem, Max = maximum, Avg = average,
Min = minimum, Med = median, C = component frequency,
CV = component-version frequency.
\end{table*}

\subsection{Tool Access and Execution Setup}
\label{subsec:tool-access-execution}
\label{subsec:api-access}
\label{subsec:tool-configuration}

To ensure comparability across heterogeneous tools, each
evaluated system is accessed through a dedicated adapter that
maps the fixed evaluation input contract to the corresponding
tool-specific access mechanism. Adapters perform three tasks:
preparation of tool-specific inputs, invocation of tool APIs
or command-line interfaces, and normalization of returned
findings into the common representation \((e,c,v',u')\).

Dependency-Track, Snyk, and Trivy are evaluated using
deterministically generated CycloneDX SBOMs derived from the
same fixed dataset contents. OSS Index and the GitHub
Advisory Database are queried directly through their
respective APIs using package and version information. Across
all tools, normalization maps returned findings to a common
representation that can be evaluated using the same matching
semantics. Ground-truth generation is based on the current
OSV API (\texttt{v1}) and its batched query interface.

Table~\ref{tab:tool-config} summarizes the concrete execution
setup used in the experiments. Where possible, configuration
details were derived directly from the implemented adapters
and orchestration scripts. For Dependency-Track, the server
configuration additionally reflects the enabled analyzers,
vulnerability sources, and package repositories used during
the experiments.

\begin{table*}[t]
\centering
\caption{Concrete tool configuration used in the experiments.}
\label{tab:tool-config}
\footnotesize
\setlength{\tabcolsep}{3.5pt}
\renewcommand{\arraystretch}{1.12}

\begin{tabular}{
>{\raggedright\arraybackslash}p{0.14\textwidth}
>{\raggedright\arraybackslash}p{0.08\textwidth}
>{\raggedright\arraybackslash}p{0.10\textwidth}
>{\raggedright\arraybackslash}p{0.12\textwidth}
>{\raggedright\arraybackslash}p{0.16\textwidth}
>{\raggedright\arraybackslash}p{0.28\textwidth}
}
\toprule
Tool & Version & Access Mode & Input Format & Execution Context & Relevant Configuration \\
\midrule

Dependency-Track &
\texttt{4.14.3} &
REST API &
CycloneDX SBOM &
Own server instance &
Enabled analyzers: Internal (without fuzzy matching), Sonatype OSS Index, and VulnDB; vulnerability sources: NVD and GitHub Advisory; repositories: PyPI, npm, NuGet, and Maven \\

Snyk &
\texttt{1.1301.2} &
CLI / backend API &
CycloneDX SBOM &
Local CLI with remote backend &
SBOM-based workflow via Bash wrapper; fixed SBOM input; default retry policy with up to 3 attempts; default timeout: 180\,s; identifier-based normalization (\texttt{CVE}/\texttt{GHSA}) \\

OSS Index &
\texttt{v3} &
REST API &
Package coordinates &
Authenticated API access &
Endpoint: \texttt{/api/v3/component-report}; batch size: 128; authentication via username/token; retry and backoff handling for transient failures and rate limits \\

GitHub Advisory Database &
Current GraphQL schema &
GraphQL API &
Package name + version &
GitHub GraphQL API &
One lookup per ground-truth component-version tuple; package-level advisory query; strict version-range evaluation without fuzzy matching; identifier-based normalization (\texttt{CVE}/\texttt{GHSA}) \\

Trivy &
\texttt{v0.69.1} &
CLI &
CycloneDX SBOM &
Local execution &
Command pattern: \texttt{trivy sbom --format json <sbom>}; fixed SBOM input; JSON output; last vulnerability DB update: 28.03.2026, 19:15:04; identifier-based normalization (\texttt{CVE}/\texttt{GHSA}); affected range approximated from \texttt{FixedVersion} when available \\
\bottomrule
\end{tabular}
\end{table*}

OWASP Dependency-Track is accessed via its REST API. Because
the platform does not support individual component queries in
the evaluation setup, the dataset is converted into a
CycloneDX SBOM and uploaded to a project-specific server-side
analysis context. Findings are retrieved via the project and
finding endpoints and normalized on the basis of component
PURLs and identifier-bearing vulnerabilities.

Snyk is accessed through its command-line interface using an
SBOM-based workflow. The generated CycloneDX SBOM is submitted
through a Bash-wrapped CLI invocation, and the returned JSON
output is parsed and normalized by the adapter. The adapter
uses a fixed retry policy and timeout to stabilize execution
under the temporally controlled workflow.

Sonatype OSS Index provides a REST API supporting batched
queries over multiple component coordinates. All dataset
components are transformed into package URLs and submitted in
component-report batches. Returned vulnerability reports are
mapped directly to the evaluated component-version tuples and
deduplicated at canonical identifier level.

The GitHub Advisory Database is accessed through the GitHub
GraphQL API. The adapter iterates over the ground-truth
component-version tuples, queries package-level advisories,
and evaluates whether the reported vulnerable version range
covers the analyzed version. This version handling is strict
and does not rely on fuzzy matching.

Trivy is executed locally through its command-line interface.
For the evaluation, the generated SBOM is analyzed using the
dependency scanner in JSON mode. Returned findings are mapped
to ecosystems via package URLs and normalized using CVE or
GHSA identifiers. If available, \texttt{FixedVersion} is used
to derive a conservative affected-version indication.

Dependency-Track, Snyk, and Trivy therefore operate on
identical CycloneDX SBOM inputs, allowing differences in
reported vulnerabilities to be attributed primarily to
tool-specific analysis and matching behavior rather than to
differences in dependency discovery.

\subsection{Results on the Ground-Truth Dataset}
\label{subsec:exp-results}

Table~\ref{tab:exp-core-results} summarizes the per-ecosystem
evaluation results on the ground-truth dataset for Snyk,
OSS Index, the GitHub Advisory Database, Trivy, and OWASP
Dependency-Track. The table reports true positives (TP),
ground-truth-relative false positives (\(FP_{GT}\)), false
negatives (FN), recall, and overlap for each ecosystem as
well as aggregated TOTAL rows.

Overall, Trivy achieves the strongest recall-oriented
performance on the current dataset. It attains the highest
total recall (\(0.96\)) while maintaining a comparatively
strong total overlap (\(0.78\)). The GitHub Advisory
Database reaches nearly the same total recall (\(0.95\)),
but at the cost of substantially more
ground-truth-relative false positives, which reduces its
total overlap to \(0.54\). OWASP
Dependency-Track (\(0.91\)) and Snyk (\(0.90\)) form a
second group with strong overall recall, whereas OSS Index
shows the weakest total recall (\(0.61\)), despite a
comparatively high total overlap (\(0.80\)). Thus, the
tools differ not only in completeness of detection but also
in the trade-off between broad recall and selective
reporting.

\begin{figure}[t]
    \centering
    \includegraphics[width=0.9\linewidth]{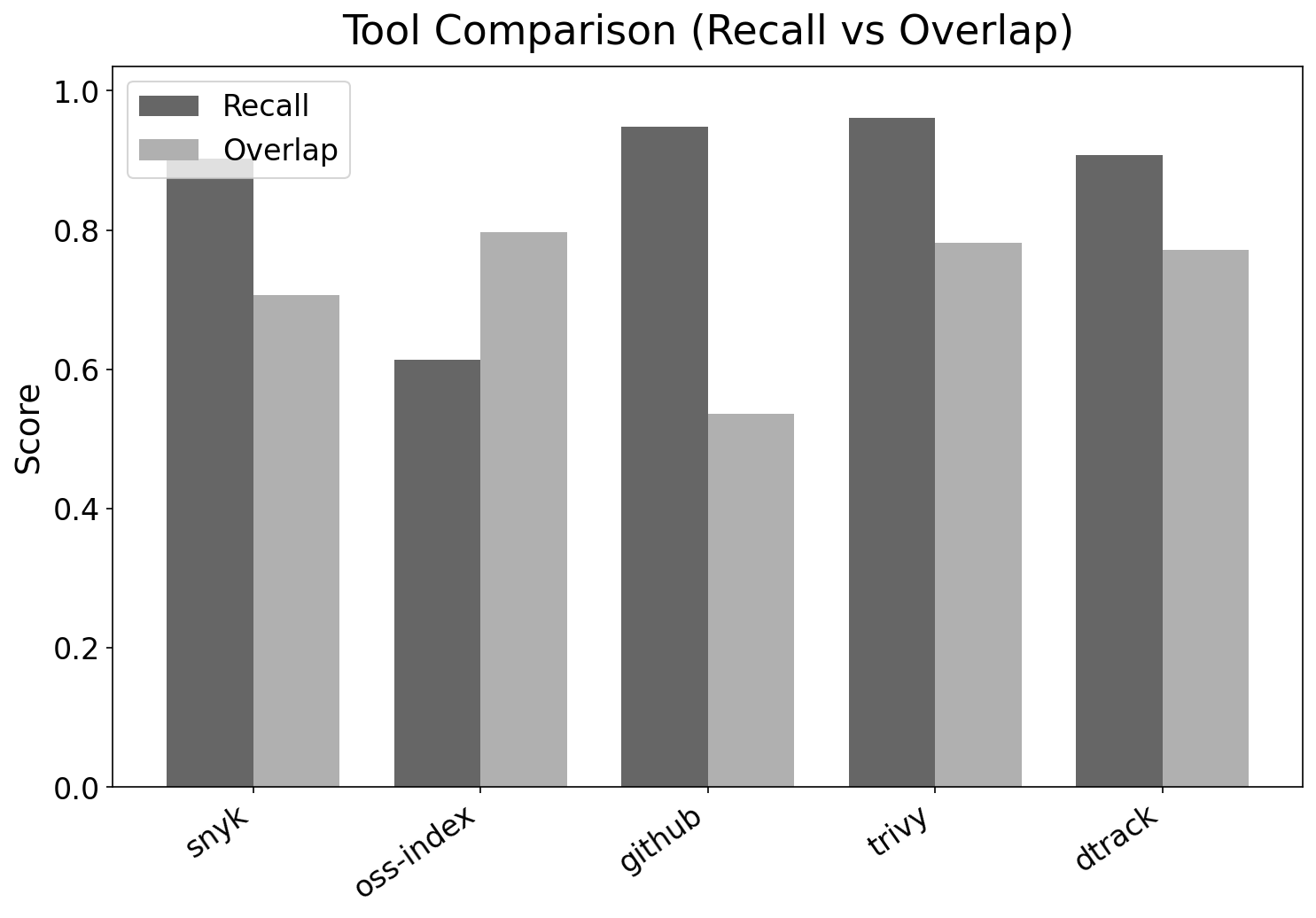}
    \caption{Comparison of the evaluated tools with respect to mean recall and mean overlap. Recall measures how completely a tool detects the ground-truth vulnerabilities, whereas overlap reflects the proportion of reported findings that correspond to true positives. The figure shows that Trivy achieves the highest overall recall, while OSS Index attains comparatively lower recall but higher overlap than some other tools.}
    \label{fig:tool-comparison}
\end{figure}

Figure~\ref{fig:tool-comparison} visualizes this trade-off
at the tool level. Trivy combines the best overall recall
with strong overlap, while the GitHub Advisory Database
emphasizes recall more strongly than reporting precision.
OSS Index appears comparatively selective once it reports a
result, but misses substantially more ground-truth
vulnerabilities overall. OWASP Dependency-Track and Snyk
occupy an intermediate position: both remain competitive in
recall, but neither dominates across all ecosystems.

At the ecosystem level, the largest spread is observed for
\texttt{Maven} and \texttt{PyPI}. For \texttt{Maven},
OWASP Dependency-Track reaches perfect recall (\(1.00\)),
followed closely by Trivy (\(0.96\)) and GitHub (\(0.93\)),
whereas Snyk (\(0.82\)) and especially OSS Index (\(0.54\))
lag behind. For \texttt{npm}, Trivy and
Dependency-Track again achieve perfect recall, GitHub and
Snyk remain strong, and OSS Index performs substantially
worse. For \texttt{NuGet}, all tools perform well, with
Snyk and GitHub reaching perfect recall and Trivy and
Dependency-Track remaining near-perfect while also
maintaining very high overlap. For \texttt{PyPI}, Trivy
(\(0.94\)) and Snyk (\(0.95\)) remain strong, GitHub
reaches \(0.86\), whereas Dependency-Track (\(0.69\)) and
OSS Index (\(0.78\)) lose substantial recall.

\begin{table*}[!t]
\centering
\caption{Per-ecosystem evaluation results on the ground-truth dataset (430 components, 1000 OSV vulnerability entries, and per-ecosystem CVE counts as shown in the table).}
\label{tab:exp-core-results}
\small
\setlength{\tabcolsep}{4pt}
\renewcommand{\arraystretch}{1.1}
\begin{tabular}{lrrrrrrrr}
\toprule
Ecosystem & Components & Vulnerabilities & CVEs & TP & FP$_{GT}$ & FN & Recall & Overlap \\
\midrule
\multicolumn{9}{c}{\textbf{snyk}} \\
\midrule
Maven & 99 & 250 & 42 & 204 & 314 & 46 & 0.82 & 0.39 \\
npm & 66 & 250 & 19 & 211 & 60 & 39 & 0.84 & 0.78 \\
NuGet & 189 & 250 & 36 & 250 & 3 & 0 & 1.00 & 0.99 \\
PyPI & 76 & 250 & 92 & 237 & 119 & 13 & 0.95 & 0.67 \\
\midrule
\textbf{TOTAL} & 430 & 1000 & 189 & 902 & 496 & 98 & 0.90 & 0.71 \\
\midrule
\multicolumn{9}{c}{\textbf{oss-index}} \\
\midrule
Maven & 99 & 250 & 42 & 136 & 52 & 114 & 0.54 & 0.72 \\
npm & 66 & 250 & 19 & 79 & 1 & 171 & 0.32 & 0.99 \\
NuGet & 189 & 250 & 36 & 204 & 12 & 46 & 0.82 & 0.94 \\
PyPI & 76 & 250 & 92 & 195 & 172 & 55 & 0.78 & 0.53 \\
\midrule
\textbf{TOTAL} & 430 & 1000 & 189 & 614 & 237 & 386 & 0.61 & 0.80 \\
\midrule
\multicolumn{9}{c}{\textbf{github}} \\
\midrule
Maven & 99 & 250 & 42 & 233 & 375 & 17 & 0.93 & 0.38 \\
npm & 66 & 250 & 19 & 250 & 133 & 0 & 1.00 & 0.65 \\
NuGet & 189 & 250 & 36 & 250 & 103 & 0 & 1.00 & 0.71 \\
PyPI & 76 & 250 & 92 & 215 & 320 & 35 & 0.86 & 0.40 \\
\midrule
\textbf{TOTAL} & 430 & 1000 & 189 & 948 & 931 & 52 & 0.95 & 0.54 \\
\midrule
\multicolumn{9}{c}{\textbf{trivy}} \\
\midrule
Maven & 99 & 250 & 42 & 240 & 193 & 10 & 0.96 & 0.55 \\
npm & 66 & 250 & 19 & 250 & 40 & 0 & 1.00 & 0.86 \\
NuGet & 189 & 250 & 36 & 237 & 3 & 13 & 0.95 & 0.99 \\
PyPI & 76 & 250 & 92 & 234 & 90 & 16 & 0.94 & 0.72 \\
\midrule
\textbf{TOTAL} & 430 & 1000 & 189 & 961 & 326 & 39 & 0.96 & 0.78 \\
\midrule
\multicolumn{9}{c}{\textbf{dtrack}} \\
\midrule
Maven & 99 & 250 & 42 & 250 & 193 & 0 & 1.00 & 0.56 \\
npm & 66 & 250 & 19 & 250 & 40 & 0 & 1.00 & 0.86 \\
NuGet & 189 & 250 & 36 & 235 & 0 & 15 & 0.94 & 1.00 \\
PyPI & 76 & 250 & 92 & 172 & 89 & 78 & 0.69 & 0.66 \\
\midrule
\textbf{TOTAL} & 430 & 1000 & 189 & 907 & 322 & 93 & 0.91 & 0.77 \\
\midrule
\bottomrule
\end{tabular}

\smallskip
\footnotesize
Recall is computed as $\mathrm{TP}/(\mathrm{TP}+\mathrm{FN})$.
Overlap is computed as $\mathrm{TP}/(\mathrm{TP}+\mathrm{FP}_{GT})$.
Here, $FP_{GT}$ denotes ground-truth-relative false positives, i.e., reported findings that are not contained in the current ground-truth snapshot.
\end{table*}

\subsection{Statistical Significance of Recall Differences}
\label{subsec:recall-significance}

To assess whether observed differences in recall between tools
are statistically significant, we model tool outcomes as
paired binary observations over the same set of ground-truth
instances. For each tool \(t\) and ground-truth instance
\(g \in GT\), we define
\begin{equation}
x_{t,g} =
\begin{cases}
1, & \text{if tool } t \text{ correctly detects } g, \\
0, & \text{otherwise.}
\end{cases}
\label{eq:binary-detection}
\end{equation}

Because all tools are evaluated on the same ground-truth
instances, the resulting observations are paired rather than
independent. The two repeated executions yielded identical
tool outputs, and the accepted run preserved the same
ground-truth snapshot throughout the controlled execution
window. Significance testing is therefore performed on the
resulting paired detection matrix over the accepted
ground-truth instances.

To test whether recall differs across multiple tools, we use
Cochran's \(Q\) test on the resulting binary detection matrix.
The null hypothesis states that all tools have the same
probability of correctly detecting a ground-truth instance.
If Cochran's \(Q\) test indicates a significant overall
difference, we perform pairwise post-hoc comparisons using the
exact McNemar test. For two tools \(A\) and \(B\), the exact McNemar test relies
on the discordant counts
\begin{align}
n_{10}^{A,B} &= |\{g \in GT \mid x_{A,g}=1,\ x_{B,g}=0\}| \label{eq:n10} \\
n_{01}^{A,B} &= |\{g \in GT \mid x_{A,g}=0,\ x_{B,g}=1\}| \label{eq:n01}
\end{align}

These counts represent ground-truth instances detected by one
tool but missed by the other. To control the family-wise error
rate across multiple pairwise comparisons, \(p\)-values are
adjusted using the Holm-Bonferroni procedure.

We do not report analogous significance tests for overlap.
Recall is defined over a common and finite set of paired
ground-truth instances, which makes Cochran's \(Q\) and
McNemar-based comparisons directly applicable. Overlap,
however, depends on \(FP_{GT}\), that is, on tool-specific
reported findings that are not contained in the current
ground-truth snapshot. These findings do not induce a shared
paired sample space across tools in the same way as
ground-truth instances do. Consequently, the statistical
framework used here for recall cannot be transferred directly
to overlap without introducing an additional canonical
universe of comparable non-ground-truth findings, which is
beyond the scope of this study.

\begin{figure}[t]
    \centering
    \includegraphics[width=0.92\linewidth]{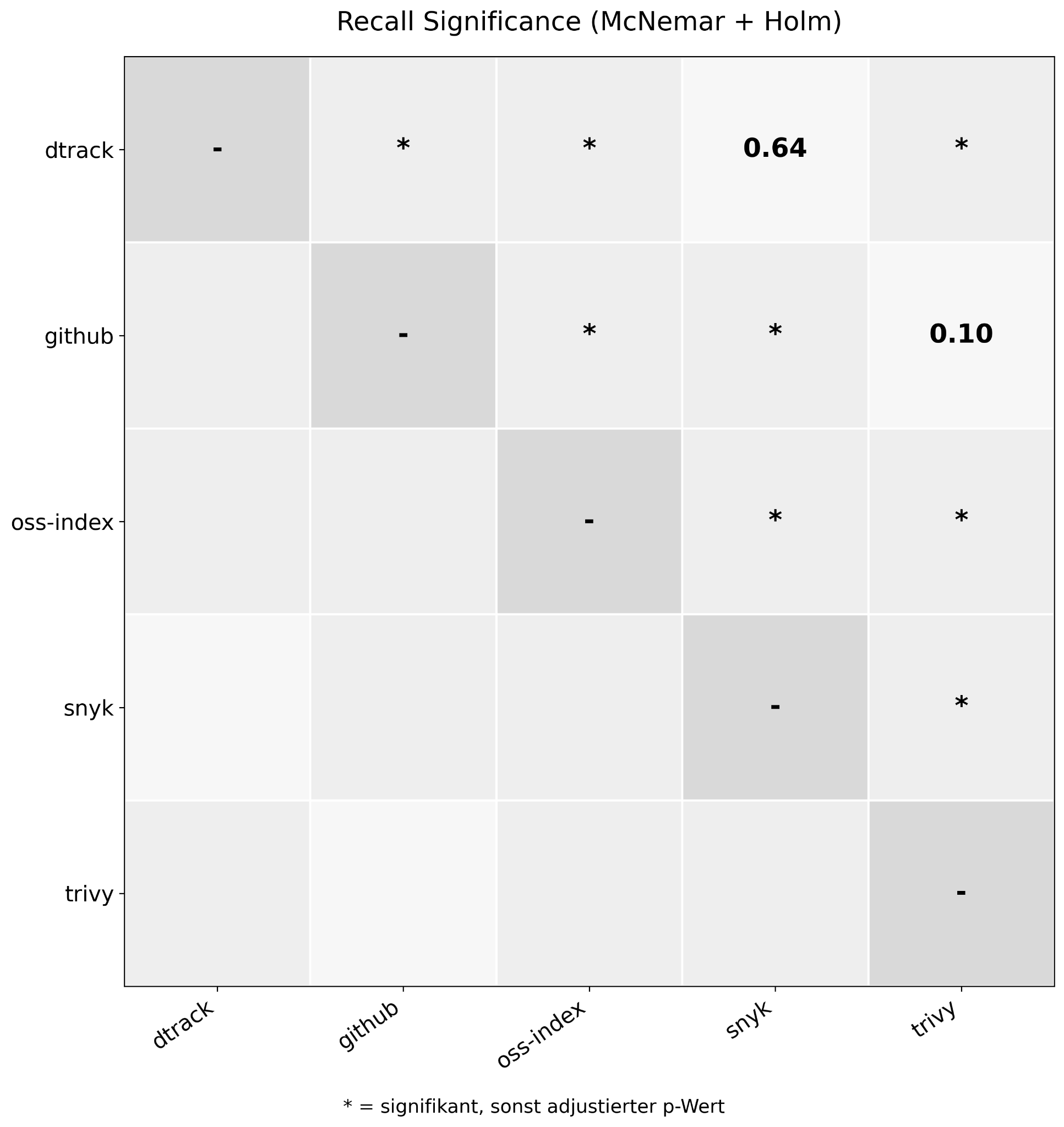}
    \caption{Significance matrix for pairwise recall comparisons.
    Each cell summarizes the post-hoc exact McNemar test for the
    corresponding tool pair after Holm correction. Significant
    differences are observed for all pairs except GitHub vs.\
    Trivy and OWASP Dependency-Track vs.\ Snyk.}
    \label{fig:significance-matrix}
\end{figure}

The global Cochran's \(Q\) test indicates a highly significant
overall difference in recall
(\(Q = 1452.81\), \(p < 0.001\)).
We then perform pairwise post-hoc comparisons using the exact
McNemar test with Holm correction. Table~\ref{tab:recall-significance}
reports the discordant counts \(n_{10}\) and \(n_{01}\), the
raw \(p\)-values, and the adjusted \(p_{\mathrm{adj}}\)-values
for all tool pairs, while
Figure~\ref{fig:significance-matrix} visualizes the
same results as a significance matrix. After Holm correction,
all pairwise differences are significant except for
GitHub vs.~Trivy (\(p_{\mathrm{adj}} = 0.104\)) and
OWASP Dependency-Track vs.~Snyk (\(p_{\mathrm{adj}} = 0.642\)).
Thus, GitHub and Trivy form the strongest recall group
without a statistically reliable difference between them,
OWASP Dependency-Track and Snyk form a statistically
indistinguishable middle group, and OSS Index is
significantly weaker than all other tools.

\begin{table*}[!t]
\centering
\caption{Pairwise significance tests for recall differences.
For each tool pair, \(n_{10}\) denotes the number of
ground-truth instances detected by Tool A but missed by Tool B,
whereas \(n_{01}\) denotes the converse case.}
\label{tab:recall-significance}
\small
\setlength{\tabcolsep}{5pt}
\renewcommand{\arraystretch}{1.10}
\begin{tabular*}{0.82\textwidth}{@{\extracolsep{\fill}}llrrrr@{}}
\toprule
Tool A & Tool B & $n_{10}$ & $n_{01}$ & $p$ & $p_{\mathrm{adj}}$ \\
\midrule
oss-index & trivy & 12 & 706 & $<0.001$ & $<0.001$ \\
github & oss-index & 730 & 62 & $<0.001$ & $<0.001$ \\
oss-index & snyk & 58 & 634 & $<0.001$ & $<0.001$ \\
dtrack & oss-index & 686 & 100 & $<0.001$ & $<0.001$ \\
dtrack & trivy & 32 & 140 & $<0.001$ & $<0.001$ \\
snyk & trivy & 72 & 190 & $<0.001$ & $<0.001$ \\
github & snyk & 194 & 102 & $<0.001$ & $<0.001$ \\
dtrack & github & 94 & 176 & $<0.001$ & $<0.001$ \\
github & trivy & 70 & 96 & 0.052 & 0.104 \\
dtrack & snyk & 192 & 182 & 0.642 & 0.642 \\
\bottomrule
\end{tabular*}
\end{table*}

\section{Discussion}
\label{sec:discussion}
\label{subsec:discussion}

The results show that SCA tool quality cannot be reduced to a
single global score. Recall, overlap, ecosystem-specific
behavior, and the statistical distinguishability of paired
detection outcomes capture different aspects of practical
usefulness. In particular, the current dataset reveals a clear
separation between tools that maximize coverage and tools that
remain more selective in their reporting.

\subsection{Interpretation of the Main Findings}

A first important observation concerns the trade-off between
recall and overlap. Trivy provides the strongest overall
balance on the present dataset: it combines the highest total
recall with a comparatively strong overlap. The GitHub
Advisory Database reaches nearly the same recall level, but
does so with a substantially larger
ground-truth-relative false-positive burden. This suggests
that GitHub is attractive when maximum coverage is the
dominant objective, whereas Trivy is more suitable when high
coverage must be achieved without sacrificing too much
selectivity.

OWASP Dependency-Track and Snyk form a second group with
strong overall recall but less favorable behavior than Trivy
in the joint consideration of coverage and overlap. OSS Index,
by contrast, is clearly more conservative: it misses
substantially more ground-truth vulnerabilities, but the
findings it does report are comparatively concentrated around
true positives. This confirms that the evaluated tools differ
not only in completeness of detection, but also in how broadly
or selectively they report vulnerabilities under identical
inputs.

The significance analysis reinforces this interpretation. The
omnibus Cochran's \(Q\) test rejects the null hypothesis of
equal recall behavior across tools, and the pairwise McNemar
tests with Holm correction leave only two non-significant
comparisons: GitHub versus Trivy and OWASP Dependency-Track
versus Snyk. Thus, the data support a three-level structure:
GitHub and Trivy form the strongest recall group,
Dependency-Track and Snyk form a statistically
indistinguishable middle group, and OSS Index is significantly
weaker than all remaining tools. At the same time, the absence
of significance between GitHub and Trivy shows that visible
differences in descriptive recall should not automatically be
overinterpreted when the paired detection patterns remain too
similar.

From a practical perspective, the results imply different tool
choices depending on the evaluation objective. If the primary
goal is to maximize coverage of known vulnerabilities, Trivy
and GitHub are the strongest candidates. If a more favorable
balance between coverage and selectivity is required, Trivy
appears particularly attractive. If conservative reporting
with comparatively high overlap is more important than maximum
recall, OSS Index may still be useful despite its weaker
coverage. The statistical similarity between Dependency-Track
and Snyk indicates that their recall performance is
effectively interchangeable on the present dataset, even
though they differ in the detailed distribution of false
positives across ecosystems.

At the same time, these findings should be interpreted
relative to the specific ground-truth snapshot and query
time. As shown in Appendix~\ref{app:temporal-gt-eval-comparison},
even a short-term repetition of the ground-truth generation
with unchanged parameters leads to small but visible shifts in
advisory assignments and tool outcomes. The core message of
that appendix is that SCA evaluation results are not fully
time-invariant: they depend not only on the matching and
reporting behavior of the tools themselves, but also on the
temporal state of the underlying vulnerability backends and
the resulting ground-truth snapshot. Consequently, the
ranking observed here should be understood as robust for the
present evaluation setting, but not as entirely independent of
temporal dataset drift.

Overall, the findings support a multi-dimensional view of SCA
tool performance. A tool that appears strong in total recall
may still be problematic in practice if its false-positive
burden is too high, while a more selective tool may be less
useful when comprehensive vulnerability coverage is required.
The present results therefore argue against a single absolute
ranking and instead favor evaluation along several dimensions:
coverage, selectivity, ecosystem sensitivity, and statistical
robustness of the observed differences.

\subsection{Ecosystem Effects and Error Patterns}

A second important observation concerns the role of the
package ecosystem. The aggregated ecosystem-level view in
Table~\ref{tab:ecosystem-summary} shows that the difficulty of
the detection task is not uniform across ecosystems.
\texttt{NuGet} is clearly the least challenging ecosystem,
with the highest mean recall (\(0.94\)) and the highest mean
overlap (\(0.93\)). It also shows the most favorable average
error profile across tools, with \(\overline{TP}=235.2\),
\(\overline{FP}_{GT}=24.2\), and \(\overline{FN}=14.8\).
In contrast, \texttt{Maven} exhibits by far the largest
average ground-truth-relative false-positive burden
(\(\overline{FP}_{GT}=225.4\)) and the lowest mean overlap
(\(0.52\)), which indicates that precise matching is
particularly difficult in this ecosystem. \texttt{npm}
appears comparatively balanced, with mean recall and mean
overlap both at \(0.83\), although its average false-negative
burden remains non-negligible
(\(\overline{FN}=42.0\)). \texttt{PyPI} remains challenging
because comparatively strong recall is repeatedly offset by a
substantial average number of ground-truth-relative false
positives (\(\overline{FP}_{GT}=158.0\)).

\begin{table*}[!t]
\centering
\caption{Aggregated ecosystem-level summary across all evaluated tools. The table reports the mean numbers of components, vulnerabilities, true positives, ground-truth-relative false positives, and false negatives across tools, as well as the mean recall and mean overlap per ecosystem. The last row reports the arithmetic mean across the four ecosystem rows.}
\label{tab:ecosystem-summary}
\small
\begin{tabular}{lrrrrrrr}
\toprule
Ecosystem & Components & Vulnerabilities & $\overline{TP}$ & $\overline{FP}_{GT}$ & $\overline{FN}$ & Mean Recall & Mean Overlap \\
\midrule
Maven & 99 & 250 & 212.6 & 225.4 & 37.4 & 0.85 & 0.52 \\
npm & 66 & 250 & 208.0 & 54.8 & 42.0 & 0.83 & 0.83 \\
NuGet & 189 & 250 & 235.2 & 24.2 & 14.8 & 0.94 & 0.93 \\
PyPI & 76 & 250 & 210.6 & 158.0 & 39.4 & 0.84 & 0.60 \\
\midrule
\textbf{Mean} & 107.5 & 250.0 & 216.6 & 115.6 & 33.4 & 0.87 & 0.72 \\
\bottomrule
\end{tabular}
\end{table*}

This ecosystem-level perspective is important because the
global ranking is not perfectly stable across all ecosystems.
For \texttt{Maven}, OWASP Dependency-Track achieves perfect
recall, followed closely by Trivy and GitHub, while OSS Index
lags far behind. For \texttt{npm}, Trivy and
Dependency-Track again achieve perfect recall, whereas
OSS Index remains substantially weaker. In \texttt{NuGet}, all
tools perform well, which suggests that vulnerability
information is comparatively easy to recover consistently in
this ecosystem. The largest spread is observed in
\texttt{PyPI}, where Trivy and Snyk remain strong, GitHub
retains acceptable recall but suffers from low overlap, and
Dependency-Track loses substantial recall. Consequently,
conclusions based only on global totals would conceal relevant
ecosystem-specific strengths and weaknesses.

Taken together, the ecosystem-level results suggest that
tool disagreement is shaped by two recurrent effects:
differences in vulnerability coverage and differences in how
advisory scope or affected-version information is mapped to
package-version instances. Low recall points to missing
coverage relative to the OSV-derived reference set, whereas
low overlap points to broader reporting relative to that same
reference snapshot. These observations support the broader
interpretation that differences between tools are driven less
by execution logic than by data coverage, identifier
resolution, and the way affected versions are represented.
\subsection{Threats to Validity and Limitations}
\label{subsec:limitations}

Several threats to validity must be considered.

\paragraph{Construct validity}
The evaluation measures agreement with an OSV-derived ground
truth rather than ground-truth derived from independent manual
validation. As a consequence, the study primarily measures how
closely tools reproduce OSV-aligned vulnerability semantics.
A related concern is that tools directly relying on OSV data
could in principle be favored by such a reference set.
However, this potential bias is limited in the present study.
OSV is itself an aggregator that consolidates vulnerability
information from multiple upstream advisory sources rather
than an entirely independent primary database. Moreover,
among the evaluated tools, only OWASP Dependency-Track could
in principle consume OSV data directly, and this source was
not enabled in the evaluated configurArXiv - A Ground-Truth-Basesdation. Most importantly,
the evaluation is output-oriented: the decisive criterion is
whether a tool reports a vulnerability for a given component
version. If a vulnerability contained in the reference set is
not reported by a tool, it is counted as a false negative
regardless of which data sources the tool could in principle
use internally.

\paragraph{Internal validity}
Although all tools are evaluated under a fixed input contract,
adapter implementations and API-specific filtering logic may
still influence the observed results. This threat is mitigated
through deterministic input generation, uniform normalization,
and temporally controlled execution.

\paragraph{External validity}
The study is limited to four package ecosystems and to
vulnerabilities disclosed primarily since approximately 2020.
The findings therefore may not generalize to ecosystems with
substantially different advisory practices or to historical
vulnerability data with weaker metadata quality.

Beyond these validity concerns, the study has several
practical limitations. First, the OSV-based ground-truth
reflects the coverage and biases of the OSV aggregation
strategy. Second, the dataset is vulnerability-balanced but
structurally heterogeneous, which may amplify
ecosystem-specific effects. Third, the analysis is limited to
recent vulnerability disclosures and to four ecosystems.
Finally, tools that rely predominantly on CVE-linked
information may be disadvantaged relative to a ground-truth
that also contains non-CVE advisories. This reflects the
diversity of real-world vulnerability data, but it also means
that tool performance must be interpreted relative to the
scope and identifier coverage of the reference dataset.
\section{Related Work}
\label{sec:related-work}

Research on vulnerability detection and analysis spans a broad
range of topics, including vulnerable dependency prevalence,
vulnerability propagation in dependency networks, empirical
evaluation of software composition analysis (SCA) tools, and
the quality of vulnerability-to-package mappings.
Table~\ref{tab:related-work-comparison} summarizes
representative quantitative studies in this area and
highlights differences in ecosystem scope, package-version
precision, and methodological focus.

Pashchenko et al. provided an early study of vulnerable open-source dependencies in the Maven ecosystem, showing that naive dependency-level counting can substantially overestimate the number of vulnerabilities that are relevant in practice \cite{pashchenko2018vulnerabilities}. This work established the importance of dependency-aware analysis for vulnerability assessment, but was limited to Maven and primarily based on CVE/NVD-derived vulnerability data. Ponta et al. later proposed a methodology for the detection, assessment, and mitigation of vulnerabilities in open-source dependencies \cite{ponta2020detection}. Although their approach moved beyond purely identifier-based matching and further strengthened dependency-aware reasoning, it also remained confined to a single ecosystem and a curated matching pipeline.

Several studies have investigated the propagation of
vulnerabilities across dependency networks. Zerouali et al. study the impact of security vulnerabilities in the npm and RubyGems dependency networks, showing that vulnerable packages may affect large numbers of direct and transitive dependents \cite{zerouali2022impact}.
Mir et al.\ further demonstrate for Maven that transitivity
and analysis granularity strongly influence conclusions about
which projects are actually affected, with fine-grained
reachability yielding markedly lower exposure estimates than
coarse dependency-level analysis \cite{mir2023granularity}.

A complementary line of work focuses on the empirical
evaluation of SCA tools themselves. Imtiaz et al.\ compare the
vulnerability reports of nine industry SCA tools on a large
case-study application and find substantial disagreement in
reported results \cite{imtiaz2021comparative}. Zhao et al.\ present an empirical study of SCA tools on Maven-based Java projects, showing that dependency resolution and version-aware analysis significantly affect tool accuracy and comparability. In addition, they highlight that the lack of context-sensitive analysis (e.g., dependency usage or reachability) leads to substantial false positives and limits the practical usefulness of SCA results \cite{zhao2023software}. More recently, Shu et al.\ extend
this line of work by evaluating SCA tools under more
challenging scenarios and report persistent weaknesses in
current tool behavior \cite{shu2025tooltoy}.

Recent work also highlights that correct vulnerability
attribution remains difficult even before tool evaluation
begins. Wu et al.\ show that identifying the affected
libraries and even the affected ecosystems for open-source
vulnerabilities is itself a non-trivial problem, and propose a
ranking-based approach to improve this mapping
\cite{wu2024holmes}. Dietrich et al.\ identify an additional
blind spot of existing SCA approaches by showing that cloned
or shaded vulnerable code can remain undetected by
conventional dependency-based analyses \cite{dietrich2024blindspots}.

Ecosystem-specific studies further illustrate the diversity of
vulnerability characteristics. Alfadel et al.\ conduct a large-scale empirical study of security vulnerabilities in Python packages and analyze their distribution, detection, and remediation characteristics within the PyPI ecosystem \cite{alfadel2023python}.

As summarized in Table~\ref{tab:related-work-comparison},
existing studies provide important insights into
vulnerability detection and dependency security, but also
exhibit recurring limitations. Many prior studies focus on a
single ecosystem, rely on product- or identifier-centric
representations, or evaluate tools without a reproducible
ground-truth at explicit package-version granularity.
Moreover, advisories and vulnerabilities are often conflated,
which can introduce duplicated findings and ambiguous
evaluation results.

We deliberately focus on vulnerability detection and
resolution for third-party dependencies rather than
vulnerabilities in application source code, which are
addressed by a largely orthogonal body of work on static and
dynamic program analysis. In contrast to prior studies, this
work introduces a cross-ecosystem, package- and
version-specific evaluation methodology based on OSV-derived
ground-truth datasets. By explicitly distinguishing
vulnerabilities from advisories and applying uniform matching
semantics across PyPI, Maven, npm, and NuGet, our approach
addresses key limitations identified in previous research and
enables a reproducible comparison of vulnerability resolution
behavior across ecosystems.

\begin{table*}[!t]
\centering
\caption{Representative quantitative studies on vulnerability detection and analysis}
\label{tab:related-work-comparison}
\small
\setlength{\tabcolsep}{4pt}
\renewcommand{\arraystretch}{1.12}

\begin{tabular*}{\textwidth}{@{\extracolsep{\fill}}
p{0.22\textwidth}
p{0.15\textwidth}
c
p{0.36\textwidth}
c
@{}}
\toprule
Work &
Ecosystems &
Ver.-prec. &
Methodological focus &
Ref. \\
\midrule

Pashchenko et al.\ (2018) &
Maven &
Limited &
Vulnerable dependency measurement and dependency-aware counting &
\cite{pashchenko2018vulnerabilities} \\

Ponta et al.\ (2020) &
Maven &
Partial &
Dependency-aware detection, assessment, and mitigation &
\cite{ponta2020detection} \\

Imtiaz et al.\ (2021) &
App case study &
Limited &
Comparative analysis of vulnerability reporting by SCA tools &
\cite{imtiaz2021comparative} \\

Zerouali et al.\ (2022) &
npm, RubyGems &
Limited &
Vulnerability propagation in dependency networks &
\cite{zerouali2022impact} \\

Mir et al.\ (2023) &
Maven &
Yes &
Effect of transitivity and fine-grained reachability &
\cite{mir2023granularity} \\

Zhao et al.\ (2023) &
Java &
Yes &
Empirical evaluation of SCA tools with version-aware analysis &
\cite{zhao2023software} \\

Alfadel et al.\ (2023) &
PyPI &
Yes &
Large-scale ecosystem-specific vulnerability analysis &
\cite{alfadel2023python} \\

Wu et al.\ (2024) &
Mixed &
Partial &
Affected-library and ecosystem identification &
\cite{wu2024holmes} \\

Dietrich et al.\ (2024) &
Maven &
No &
Blind spots caused by cloned or shaded vulnerable code &
\cite{dietrich2024blindspots} \\

Shu et al.\ (2025) &
Mixed &
Partial &
Evaluation of SCA tools in challenging scenarios &
\cite{shu2025tooltoy} \\

\textbf{This work} &
\textbf{PyPI, Maven, npm, NuGet} &
\textbf{Yes} &
\textbf{Cross-ecosystem vulnerability resolution using OSV-derived ground-truth} &
-- \\

\bottomrule
\end{tabular*}

\smallskip
\raggedright
\small
\textit{Note:} ``Ver.-prec.'' denotes explicit resolution at the
package-version level.
\end{table*}

A key distinction from vulnerability-centric benchmarks is
that our evaluation operates on explicit package-version
instances rather than on advisory-level or identifier-level
matches.

Prior empirical studies indicate that differences between SCA
tools are strongly influenced by underlying data sources,
aggregation strategies, and version semantics rather than by
tool execution logic alone
\cite{imtiaz2021comparative,zhao2023software}.

Empirical work on dependency ecosystems further shows that
vulnerability propagation can easily be over-approximated when
tools reason over dependency graphs without sufficiently
precise package-version confirmation. In particular, studies
of npm, RubyGems, and Maven demonstrate that a small number of
vulnerable packages can affect large numbers of dependent
releases, and that transitivity and analysis granularity
substantially change conclusions about actual exposure
\cite{zerouali2022impact,mir2023granularity}.

In addition, comparative evaluations of SCA tools indicate
that advisory-level affected-version ranges are frequently
underspecified or overly conservative, such that mechanically
applying range-based matching introduces additional false
positives and ambiguous results
\cite{imtiaz2021comparative,ponta2020detection}. Recent work
also shows that even the identification of the affected
library and ecosystem for a vulnerability remains difficult,
which further complicates evaluation if package-level
semantics are not made explicit \cite{wu2024holmes}.

Together with earlier large-scale studies on vulnerable
dependencies and vulnerability propagation
\cite{pashchenko2018vulnerabilities,ponta2020detection,
zerouali2022impact,mir2023granularity}, this body of work
motivates the component-centric, version-specific evaluation
methodology adopted in this paper.

\section{Conclusions and Outlook}
\label{sec:conclusion}

This study presented a systematic, component-centric evaluation of software
composition analysis tools using an OSV-derived ground-truth dataset. By
constructing a reproducible dataset at the component--version level, the study
reduced ambiguities caused by heterogeneous advisory aggregation, inconsistent
identifier coverage, and incomplete vulnerability metadata. This enabled a
controlled comparison of tool behavior across multiple ecosystems under a
shared input contract and uniform matching semantics.

The results show substantial differences between tools in both vulnerability
coverage and reporting selectivity. These differences are not explained solely
by tool execution or interface design, but arise primarily from differences in
advisory coverage, aggregation strategies, identifier resolution, and
version-matching logic. In many cases, identical component-version inputs led
to different vulnerability assignments across tools. This indicates that
accurate vulnerability assessment is fundamentally a data and semantics
problem, not merely a tooling problem.

At the same time, the evaluation shows that tool quality cannot be reduced to
a single global score. Recall and overlap capture different aspects of
practical usefulness, and the observed ranking is not fully stable across all
ecosystems. Some tools emphasize broader vulnerability coverage, whereas
others behave more selectively and therefore achieve higher agreement with the
reference set once they report a finding. The results therefore argue for a
multi-dimensional view of SCA tool evaluation that considers coverage,
selectivity, ecosystem sensitivity, and statistical robustness jointly.

The component-version--level ground-truth constructed in this work provides a
reproducible basis for benchmarking software composition analysis tools. It
supports fair and transparent comparisons across tools and over time, and it
enables more precise analysis of advisory aggregation effects, database
coverage, and vulnerability resolution strategies. Such datasets are essential
for moving tool evaluation from anecdotal comparisons toward controlled,
empirical assessment.

The study also opens several directions for future research. First, the
constructed datasets enable longitudinal analyses in which identical
component-version sets are re-evaluated over time in order to study the
evolution of vulnerability disclosures, advisory database behavior, and tool
coverage dynamics. Second, the methodology can be extended to additional
ecosystems, historical time periods, and alternative reference sources in
order to test how robust the observed patterns remain under broader
conditions.

Another promising direction concerns ensemble- or voting-based approaches for
improving the robustness of vulnerability detection. Combining tool outputs
through explicit aggregation rules may offer a useful trade-off between
broader coverage and reduced instability relative to relying on a single tool
alone. Future work should therefore not only compare tools individually, but
also investigate whether consensus-based aggregation can serve as a more
reliable basis for practical vulnerability assessment.

A further promising direction concerns AI-assisted vulnerability analysis.
Recent work has begun to explore AI-based support for vulnerability validation
and remediation~\cite{lotfi2025llm_vuln_validation,kathi2025ai_dependency}.
However, the use of AI techniques for improving the assignment of known
vulnerabilities to open-source libraries remains comparatively underexplored,
especially in dependency analysis and tool evaluation. The findings of this
study suggest that the practical utility of such approaches depends critically
on the availability of precise, component-version--level ground-truth.
Learning-based methods are therefore best understood as complementary
techniques that may help resolve ambiguous or incomplete vulnerability
metadata, rather than as replacements for explicit and semantically grounded
vulnerability representations.

Overall, this work shows that reproducible, component-centric ground-truth is
a prerequisite for rigorous and comparable vulnerability assessment. By making
vulnerability resolution explicit, component-centric, and reproducible, the
study provides a methodological foundation on top of which both rule-based and
learning-based approaches can be evaluated systematically and fairly.

\section*{Acknowledgment}

The authors used OpenAI GPT models (GPT-5.2 to GPT-5.4) for language editing and manuscript refinement. All scientific content, analyses, interpretations, and conclusions remain the responsibility of the authors.



\bibliographystyle{IEEEtran}
\bibliography{references}

\appendices
\FloatBarrier
\section{Temporal Comparison of Two Ground-Truth Snapshots and Their Tool Evaluations}
\label{app:temporal-gt-eval-comparison}

\begin{table*}[!t]
\centering
\caption{Summary of ground-truth changes between 28.03.2026 and 10.04.2026 with visible tool effects.}
\label{tab:gt-change-summary}
\small
\setlength{\tabcolsep}{3pt}
\renewcommand{\arraystretch}{1.08}
\begin{tabular*}{\textwidth}{@{\extracolsep{\fill}}
p{0.08\textwidth}
p{0.07\textwidth}
p{0.07\textwidth}
p{0.09\textwidth}
p{0.09\textwidth}
p{0.50\textwidth}
@{}}
\toprule
Eco & Removed & Added & $\Delta$ CVE-F. & $\Delta$ CVEs & Visible tool effects \\
\midrule
Maven & 0 & 0 & 0 & 0 &
No ground-truth change. Dependency-Track, GitHub,
OSS Index, and Snyk remain stable in Maven. Trivy still
improves from \(240/193/10\) to \(250/193/0\) (TP/FP/FN),
indicating a temporal change in tool or backend data. \\

npm & 20 & 20 & 0 & 0 &
Strongest effect on the evaluation. GitHub shows
\(\Delta FP=+60\), while Dependency-Track, Snyk, and Trivy
each show \(\Delta FP=+20\). OSS Index is the only tool that
clearly benefits, with \(\Delta TP=+20\) and
\(\Delta FN=-20\). \\

NuGet & 0 & 0 & 0 & 0 &
Ground truth unchanged. Only OSS Index slightly reduces its
false positives (\(\Delta FP=-5\)); all other tools remain
stable in their core metrics. \\

PyPI & 12 & 12 & +2 & +1 &
All five tools react visibly. Dependency-Track, GitHub,
OSS Index, Snyk, and Trivy gain additional true positives
and reduce false negatives; at the same time, false
positives increase for most tools. GitHub is the only
exception, with \(\Delta FP=-16\). \\

\midrule
TOTAL & 32 & 32 & +2 & +2 &
The observed differences are explained mainly by
\texttt{npm} and \texttt{PyPI}, but not exclusively.
In particular, Trivy improves in \texttt{Maven} despite an
unchanged ground-truth, underscoring the additional role of
temporally varying tool data. \\
\bottomrule
\end{tabular*}

\par\vspace{0.6em}
\raggedright
\footnotesize
\textit{Legend:} Removed = records contained in the earlier
ground-truth snapshot but no longer contained in the later
snapshot; Added = records newly contained in the later
ground-truth snapshot. \(\Delta\) CVE-F. denotes the change
in the number of CVE-backed findings, and \(\Delta\) CVEs the
change in the number of distinct CVE identifiers. The
described tool effects summarize visible changes in the
evaluation results between both snapshots.
\end{table*}

\begin{table*}[!t]
\centering
\caption{Overall comparison of the tool evaluations between 28.03.2026 and 10.04.2026.}
\label{tab:tool-diff-overview}
\scriptsize
\setlength{\tabcolsep}{2.5pt}
\renewcommand{\arraystretch}{1.05}
\begin{tabular*}{\textwidth}{@{\extracolsep{\fill}}lrrrrrrrrrrr@{}}
\toprule
Tool & TP$_{28}$ & TP$_{10}$ & $\Delta$TP & FP$_{28}$ & FP$_{10}$ & $\Delta$FP & FN$_{28}$ & FN$_{10}$ & $\Delta$FN & $\Delta$Recall & $\Delta$Overlap \\
\midrule
Dependency-Track & 907 & 908 & +1  & 322 & 350 & +28 & 93  & 92  & -1  & +0.001 & -0.016 \\
GitHub           & 948 & 949 & +1  & 931 & 975 & +44 & 52  & 51  & -1  & +0.001 & -0.012 \\
OSS Index        & 614 & 636 & +22 & 237 & 240 & +3  & 386 & 364 & -22 & +0.022 & +0.004 \\
Snyk             & 902 & 905 & +3  & 496 & 539 & +43 & 98  & 95  & -3  & +0.003 & -0.018 \\
Trivy            & 961 & 978 & +17 & 326 & 356 & +30 & 39  & 22  & -17 & +0.017 & -0.014 \\
\bottomrule
\end{tabular*}
\end{table*}

\begin{table*}[!b]
\centering
\caption{Linkage between ground-truth changes and the observed tool deltas per ecosystem. The tool columns report $\Delta$TP/$\Delta$FP/$\Delta$FN between 28.03.2026 and 10.04.2026.}
\label{tab:gt-tool-linkage}
\scriptsize
\setlength{\tabcolsep}{3.5pt}
\renewcommand{\arraystretch}{1.08}
\begin{tabular*}{\textwidth}{@{\extracolsep{\fill}}lrrrrccccc@{}}
\toprule
Eco & Rem. & Add. & $\Delta$CVE-F. & $\Delta$CVEs &
Dependency-Track &
GitHub &
OSS Index &
Snyk &
Trivy \\
\midrule
Maven &
0 & 0 & 0 & 0 &
$0/+0/+0$ &
$0/+0/+0$ &
$0/+0/+0$ &
$0/+0/+0$ &
$+10/+0/-10$ \\

npm &
20 & 20 & 0 & 0 &
$0/+20/+0$ &
$0/+60/+0$ &
$+20/+0/-20$ &
$0/+20/+0$ &
$0/+20/+0$ \\

NuGet &
0 & 0 & 0 & 0 &
$0/+0/+0$ &
$0/+0/+0$ &
$0/-5/+0$ &
$0/+0/+0$ &
$0/+0/+0$ \\

PyPI &
12 & 12 & +2 & +1 &
$+1/+8/-1$ &
$+1/-16/-1$ &
$+2/+8/-2$ &
$+3/+23/-3$ &
$+7/+10/-7$ \\

\midrule
TOTAL &
32 & 32 & +2 & +2 &
$+1/+28/-1$ &
$+1/+44/-1$ &
$+22/+3/-22$ &
$+3/+43/-3$ &
$+17/+30/-17$ \\
\bottomrule
\end{tabular*}

\par\vspace{0.6em}
\raggedright
\footnotesize
\textit{Note:} Rem. = removed ground-truth entries; Add. = newly
added ground-truth entries. The tool columns report the
observed differences in the evaluation results between the
two time points as $\Delta$TP/$\Delta$FP/$\Delta$FN. These
differences cannot necessarily be explained monocausally by
the ground-truth changes.
\end{table*}

\begin{table*}[p]
\centering
\caption{Per-ecosystem comparison of the five evaluated tools between 28.03.2026 and 10.04.2026.}
\label{tab:tool-diff-detailed}
\scriptsize
\setlength{\tabcolsep}{3pt}
\renewcommand{\arraystretch}{1.05}
\begin{tabular*}{\textwidth}{@{\extracolsep{\fill}}llrrrrrrrrr@{}}
\toprule
Tool & Eco & TP$_{28}$ & TP$_{10}$ & $\Delta$TP & FP$_{28}$ & FP$_{10}$ & $\Delta$FP & FN$_{28}$ & FN$_{10}$ & $\Delta$FN \\
\midrule
Dependency-Track & Maven & 250 & 250 & 0  & 193 & 193 & 0  & 0  & 0  & 0 \\
                 & npm   & 250 & 250 & 0  & 40  & 60  & +20 & 0  & 0  & 0 \\
                 & NuGet & 235 & 235 & 0  & 0   & 0   & 0  & 15 & 15 & 0 \\
                 & PyPI  & 172 & 173 & +1 & 89  & 97  & +8 & 78 & 77 & -1 \\
                 & TOTAL & 907 & 908 & +1 & 322 & 350 & +28 & 93 & 92 & -1 \\
\midrule
GitHub          & Maven & 233 & 233 & 0  & 375 & 375 & 0   & 17 & 17 & 0 \\
                & npm   & 250 & 250 & 0  & 133 & 193 & +60 & 0  & 0  & 0 \\
                & NuGet & 250 & 250 & 0  & 103 & 103 & 0   & 0  & 0  & 0 \\
                & PyPI  & 215 & 216 & +1 & 320 & 304 & -16 & 35 & 34 & -1 \\
                & TOTAL & 948 & 949 & +1 & 931 & 975 & +44 & 52 & 51 & -1 \\
\midrule
OSS Index       & Maven & 136 & 136 & 0   & 52  & 52  & 0  & 114 & 114 & 0 \\
                & npm   & 79  & 99  & +20 & 1   & 1   & 0  & 171 & 151 & -20 \\
                & NuGet & 204 & 204 & 0   & 12  & 7   & -5 & 46  & 46  & 0 \\
                & PyPI  & 195 & 197 & +2  & 172 & 180 & +8 & 55  & 53  & -2 \\
                & TOTAL & 614 & 636 & +22 & 237 & 240 & +3 & 386 & 364 & -22 \\
\midrule
Snyk            & Maven & 204 & 204 & 0  & 314 & 314 & 0   & 46 & 46 & 0 \\
                & npm   & 211 & 211 & 0  & 60  & 80  & +20 & 39 & 39 & 0 \\
                & NuGet & 250 & 250 & 0  & 3   & 3   & 0   & 0  & 0  & 0 \\
                & PyPI  & 237 & 240 & +3 & 119 & 142 & +23 & 13 & 10 & -3 \\
                & TOTAL & 902 & 905 & +3 & 496 & 539 & +43 & 98 & 95 & -3 \\
\midrule
Trivy           & Maven & 240 & 250 & +10 & 193 & 193 & 0   & 10 & 0  & -10 \\
                & npm   & 250 & 250 & 0   & 40  & 60  & +20 & 0  & 0  & 0 \\
                & NuGet & 237 & 237 & 0   & 3   & 3   & 0   & 13 & 13 & 0 \\
                & PyPI  & 234 & 241 & +7  & 90  & 100 & +10 & 16 & 9  & -7 \\
                & TOTAL & 961 & 978 & +17 & 326 & 356 & +30 & 39 & 22 & -17 \\
\bottomrule
\end{tabular*}

\par\vspace{0.6em}
\raggedright
\footnotesize
\textit{Legend:} TP = true positives, FP = ground-truth-relative false positives,
FN = false negatives. The subscripts 28 and 10 denote the evaluation results
for the ground-truth snapshots from 28.03.2026 and 10.04.2026, respectively.
\(\Delta\) indicates the difference between both snapshots.
\end{table*}

To examine the temporal stability of the underlying
vulnerability data and its impact on tool assessment more
systematically, the ground-truth generation was repeated
after an interval of 13 days using unchanged generation
parameters. The purpose of this repeated run was not to
construct an alternative dataset, but to make visible how
strongly the underlying vulnerability sources can change even
over a short period of time and to what extent these changes
affect the evaluation results of the considered tools. The
comparison between the original ground-truth snapshot from
28.03.2026 and the updated snapshot from 10.04.2026 thus
serves both as a targeted stability test of the reference set
and as an illustration of the temporal sensitivity of
data-driven SCA evaluations.

The effective generation parameters remained unchanged.
Likewise, the number of components, the number of OSV
entries, and their distribution across the ecosystems
remained constant. Structural changes occurred exclusively in
the CVE-related metrics and in individual vulnerability
assignments. At the same time, the tool evaluations show that
the differences between the two time points cannot be
explained solely by changes in the ground-truth. Rather, the
observations reflect the combined effect of ground-truth
drift and temporally changing tool or backend data.

As Table~\ref{tab:gt-change-summary} shows, the structural
changes in the ground-truth affect only \texttt{npm} and
\texttt{PyPI}. In \texttt{npm}, 20 entries were removed and
20 new ones were added, without any change in the aggregated
CVE-related metrics. In \texttt{PyPI}, 12 entries were
removed and 12 new ones were added; at the same time, the
number of CVE-backed findings increases from 203 to 205 and
the number of distinct CVE identifiers from 92 to 93.
Globally, the number of CVE-backed findings therefore rises
from 924 to 926, and the number of globally unique CVEs from
188 to 190.

The provided tool evaluations react very differently to these
changes. Table~\ref{tab:tool-diff-overview} summarizes the
global differences, whereas Table~\ref{tab:gt-tool-linkage}
directly links the ground-truth changes to the observed tool
deltas at the ecosystem level. OSS Index benefits the most,
with \(+22\) true positives and \(-22\) false negatives,
which is mainly due to improvements in \texttt{npm}. Trivy
achieves the largest absolute recall gain (\(+17\) TP,
\(-17\) FN), but simultaneously loses overlap because
additional false positives occur. Snyk improves only slightly
in recall, but accumulates \(+43\) additional false
positives. GitHub keeps recall and FN almost constant, but
likewise loses overlap due to \(+44\) additional false
positives. Dependency-Track remains the most stable overall,
but also shifts toward a higher FP burden.

For the interpretation, it is particularly important that not
all tool shifts can be traced back directly to changes in the
ground-truth. The clearest evidence for this is provided by
\texttt{Maven}: the ground-truth remains unchanged there,
while Trivy simultaneously increases from 240 to 250 TP and
reduces its false negatives from 10 to 0. This difference
clearly points to a temporal change in the underlying tool
data or resolver logic. By contrast, the changes in
\texttt{npm} and \texttt{PyPI} fit the observed
ground-truth adjustments much more closely: \texttt{npm}
mainly drives FP increases for GitHub, Dependency-Track,
Snyk, and Trivy, as well as a pronounced recall gain for OSS
Index; \texttt{PyPI} explains the largest share of the
additional TPs for Snyk and Trivy.

Table~\ref{tab:gt-tool-linkage} makes this relation explicit
by linking structural ground-truth changes to the observed
tool deltas at ecosystem level. The overall pattern suggests
that the observed differences should not be understood as the
effect of a simple dataset replacement. At the structural
level, the ground-truth remains largely stable: the total
number of components, the total number of OSV entries, and
their distribution across the four ecosystems do not change.
What does change, however, are individual advisory
assignments, especially in \texttt{npm} and \texttt{PyPI}. In
\texttt{npm}, the changes are almost entirely driven by a
systematic exchange of \texttt{vite}-related advisory
records, whereas in \texttt{PyPI} they are distributed across
multiple packages and include both substitutions and pure
additions or removals. Accordingly, the tool results do not
shift in a uniform way either: some tools gain recall,
others mainly accumulate additional false positives, and
still others show both effects simultaneously.

It is important to note that the changed identifiers should
not be understood as mere renamings. In the OSV data model,
vulnerability records are aggregated from multiple upstream
sources and linked to source-specific advisory IDs, CVE
aliases, and package-specific affected-version ranges. If
upstream advisories, alias relationships, or
affected-version information change, the same package version
may in a later snapshot be represented by a different
advisory record or by a different advisory/alias
combination. The differences visible in
Tables~\ref{tab:gt-diff-npm} and \ref{tab:gt-diff-PyPI}
should therefore be interpreted more appropriately as
replacements or reassignments of advisory records rather than
as simple renamings.

Part of these differences can additionally be explained by
the selection logic of the ground-truth pipeline. Across all
collectors, generation does not begin with a complete
representation of the respective ecosystem, but with a
curated package set. This is followed by an ecosystem-
specific reduction of the search space: only a subset of the
available versions is processed further, partly under
consideration of the publication date and partly on the basis
of a fixed selection along the version history. For each
selected package version, an OSV query is then performed,
from whose results only a limited subset of advisory records
is admitted to further processing; duplicate advisory IDs are
deduplicated within the same package version. In addition,
collection may stop early as soon as the target size for an
ecosystem has been reached. For \texttt{NuGet}, this is
further refined by a distributed selection across the
remaining versions after date filtering rather than by taking
a simple prefix of the version history. As a result, the
final ground-truth in all considered ecosystems is the result
of multiple consecutive selection steps rather than a simple
1:1 representation of the current OSV inventory.

Against this background, it is plausible that newly added
advisories in a later run enter the final selection and
displace older entries, even though these continue to exist
in OSV. Especially for package versions with many associated
vulnerability records, the composition of the dataset can
therefore change without any earlier entry having been
deleted or invalidated. \emph{Removed} entries should thus be
understood methodologically as records that were no longer
included in the final ground-truth in the later snapshot.
This effect can be triggered by changes in the OSV inventory
itself, by a changed composition or ordering of the returned
advisories, and by the downstream balancing and
target-size mechanisms of the pipeline. The statistics files
additionally support this interpretation, because they
explicitly describe the final reference set as the result of
a \emph{post-balancing dataset} or a balanced final dataset.

A manual plausibility check against OSV supports this
interpretation. The affected advisory families listed in
Tables~\ref{tab:gt-diff-npm} and \ref{tab:gt-diff-PyPI} can
be verified in OSV; this applies both to the earlier and to
the later \texttt{vite}-related entries in \texttt{npm}, as
well as to the newly appearing \texttt{PyPI} advisories, for
example for \texttt{cryptography}. Moreover, OSV explicitly
documents for \texttt{PYSEC-2023-175} that the earlier
mapping to \texttt{CVE-2023-4863} was replaced by a refined
mapping to \texttt{CVE-2023-5129}. The tables therefore do
not capture mere format changes, but actual changes in the
advisory-based resolution and alias interpretation.

Overall, this observation strengthens the central
interpretation of this appendix section. The temporally
shifted comparison not only reveals a limited instability of
the reference set, but also shows that the evaluation results
depend both on the temporal state of the ground-truth and on
the temporal development of the queried vulnerability
backends. This is also supported by the detailed
per-ecosystem comparison across all tools in
Table~\ref{tab:tool-diff-detailed}: while the structural
ground-truth changes are concentrated in \texttt{npm} and
\texttt{PyPI}, individual tools simultaneously change in
ecosystems whose ground-truth structure remains unchanged.
The comparison therefore points to an additional effect of
time-dependent backend drift alongside pure ground-truth
drift.

Table~\ref{tab:tool-diff-detailed} reports the detailed
per-ecosystem comparison for all five tools in a unified
view. The detailed ground-truth differences are presented in
Table~\ref{tab:gt-diff-npm} for \texttt{npm} and in
Table~\ref{tab:gt-diff-PyPI} for \texttt{PyPI}. For
\texttt{npm}, there is a consistent pairwise exchange of the
same \texttt{vite}-related vulnerability type across 20
versions. In \texttt{PyPI}, by contrast, the changes are
distributed across several packages and include both
individual substitutions and pure additions or removals.

Overall, the comparison shows that even small temporal
changes in the ground-truth can visibly shift the result
profiles of the tools, despite unchanged dataset size and an
identical sampling configuration. More importantly, however,
these shifts do not originate solely from the reference side,
but likewise from the temporal evolution of the advisory and
vulnerability backends used. The assessment of a tool should
therefore not be interpreted as a fully time-invariant
property, but always relative to the respective
ground-truth snapshot and query time.

\begin{table*}[p]
\centering
\caption{Pairwise ground-truth changes in \texttt{npm}. For all affected \texttt{vite} versions, the same earlier entry was replaced by the same new entry.}
\label{tab:gt-diff-npm}
\scriptsize
\setlength{\tabcolsep}{4pt}
\renewcommand{\arraystretch}{1.05}
\begin{tabular*}{\textwidth}{@{\extracolsep{\fill}}llllll@{}}
\toprule
Version & Removed ID & Removed CVE & Added ID & Added CVE & Component \\
\midrule
0.1.0  & GHSA-vg6x-rcgg-rjx6 & CVE-2025-24010 & GHSA-4w7w-66w2-5vf9 & CVE-2026-39365 & vite \\
0.1.1  & GHSA-vg6x-rcgg-rjx6 & CVE-2025-24010 & GHSA-4w7w-66w2-5vf9 & CVE-2026-39365 & vite \\
0.1.2  & GHSA-vg6x-rcgg-rjx6 & CVE-2025-24010 & GHSA-4w7w-66w2-5vf9 & CVE-2026-39365 & vite \\
0.2.0  & GHSA-vg6x-rcgg-rjx6 & CVE-2025-24010 & GHSA-4w7w-66w2-5vf9 & CVE-2026-39365 & vite \\
0.3.0  & GHSA-vg6x-rcgg-rjx6 & CVE-2025-24010 & GHSA-4w7w-66w2-5vf9 & CVE-2026-39365 & vite \\
0.3.1  & GHSA-vg6x-rcgg-rjx6 & CVE-2025-24010 & GHSA-4w7w-66w2-5vf9 & CVE-2026-39365 & vite \\
0.3.2  & GHSA-vg6x-rcgg-rjx6 & CVE-2025-24010 & GHSA-4w7w-66w2-5vf9 & CVE-2026-39365 & vite \\
0.4.0  & GHSA-vg6x-rcgg-rjx6 & CVE-2025-24010 & GHSA-4w7w-66w2-5vf9 & CVE-2026-39365 & vite \\
0.5.0  & GHSA-vg6x-rcgg-rjx6 & CVE-2025-24010 & GHSA-4w7w-66w2-5vf9 & CVE-2026-39365 & vite \\
0.5.1  & GHSA-vg6x-rcgg-rjx6 & CVE-2025-24010 & GHSA-4w7w-66w2-5vf9 & CVE-2026-39365 & vite \\
0.5.2  & GHSA-vg6x-rcgg-rjx6 & CVE-2025-24010 & GHSA-4w7w-66w2-5vf9 & CVE-2026-39365 & vite \\
0.5.3  & GHSA-vg6x-rcgg-rjx6 & CVE-2025-24010 & GHSA-4w7w-66w2-5vf9 & CVE-2026-39365 & vite \\
0.6.0  & GHSA-vg6x-rcgg-rjx6 & CVE-2025-24010 & GHSA-4w7w-66w2-5vf9 & CVE-2026-39365 & vite \\
0.6.1  & GHSA-vg6x-rcgg-rjx6 & CVE-2025-24010 & GHSA-4w7w-66w2-5vf9 & CVE-2026-39365 & vite \\
0.7.0  & GHSA-vg6x-rcgg-rjx6 & CVE-2025-24010 & GHSA-4w7w-66w2-5vf9 & CVE-2026-39365 & vite \\
0.8.0  & GHSA-vg6x-rcgg-rjx6 & CVE-2025-24010 & GHSA-4w7w-66w2-5vf9 & CVE-2026-39365 & vite \\
0.8.1  & GHSA-vg6x-rcgg-rjx6 & CVE-2025-24010 & GHSA-4w7w-66w2-5vf9 & CVE-2026-39365 & vite \\
0.9.0  & GHSA-vg6x-rcgg-rjx6 & CVE-2025-24010 & GHSA-4w7w-66w2-5vf9 & CVE-2026-39365 & vite \\
0.9.1  & GHSA-vg6x-rcgg-rjx6 & CVE-2025-24010 & GHSA-4w7w-66w2-5vf9 & CVE-2026-39365 & vite \\
0.10.0 & GHSA-vg6x-rcgg-rjx6 & CVE-2025-24010 & GHSA-4w7w-66w2-5vf9 & CVE-2026-39365 & vite \\
\bottomrule
\end{tabular*}

\par\vspace{0.6em}
\raggedright
\footnotesize
\textit{Note:} ``Removed'' and ``Added'' refer only to differences between
the two generated ground-truth snapshots, not to deletion from or insertion
into OSV or other upstream vulnerability databases.
\end{table*}

\begin{table*}[p]
\centering
\caption{Ground-truth changes in \texttt{PyPI}, grouped by component.}
\label{tab:gt-diff-PyPI}
\scriptsize
\setlength{\tabcolsep}{4pt}
\renewcommand{\arraystretch}{1.05}
\begin{tabular*}{\textwidth}{@{\extracolsep{\fill}}
p{0.19\textwidth}
p{0.08\textwidth}
p{0.17\textwidth}
p{0.10\textwidth}
p{0.24\textwidth}
p{0.14\textwidth}
@{}}
\toprule
Change & Eco & Component & Version & Vulnerability-ID & CVE \\
\midrule

\multicolumn{6}{l}{\textit{aiohttp}} \\
removed & PyPI & aiohttp & 3.9.5 & GHSA-fh55-r93g-j68g & CVE-2025-69230 \\
removed & PyPI & aiohttp & 3.9.5 & GHSA-g84x-mcqj-x9qq & CVE-2025-69229 \\
removed & PyPI & aiohttp & 3.9.5 & GHSA-jj3x-wxrx-4x23 & CVE-2025-69227 \\
removed & PyPI & aiohttp & 3.9.5 & GHSA-mqqc-3gqh-h2x8 & CVE-2025-69225 \\
added & PyPI & aiohttp & 3.9.5 & GHSA-2vrm-gr82-f7m5 & CVE-2026-34514 \\
added & PyPI & aiohttp & 3.9.5 & GHSA-3wq7-rqq7-wx6j & CVE-2026-34517 \\
added & PyPI & aiohttp & 3.9.5 & GHSA-63hf-3vf5-4wqf & CVE-2026-34520 \\
added & PyPI & aiohttp & 3.9.5 & GHSA-966j-vmvw-g2g9 & CVE-2026-34518 \\

\multicolumn{6}{l}{\textit{cryptography}} \\
removed & PyPI & cryptography & 46.0.4 & GHSA-m959-cc7f-wv43 & CVE-2026-34073 \\
removed & PyPI & cryptography & 46.0.4 & GHSA-r6ph-v2qm-q3c2 & CVE-2026-26007 \\
added & PyPI & cryptography & 46.0.5 & GHSA-p423-j2cm-9vmq & CVE-2026-39892 \\
added & PyPI & cryptography & 46.0.6 & GHSA-p423-j2cm-9vmq & CVE-2026-39892 \\

\multicolumn{6}{l}{\textit{jwcrypto}} \\
removed & PyPI & jwcrypto & 1.5.4 & GHSA-j857-7rvv-vj97 & CVE-2024-28102 \\
added & PyPI & jwcrypto & 1.5.5 & GHSA-fjrm-76x2-c4q4 & CVE-2026-39373 \\
added & PyPI & jwcrypto & 1.5.6 & GHSA-fjrm-76x2-c4q4 & CVE-2026-39373 \\

\multicolumn{6}{l}{\textit{pillow}} \\
removed & PyPI & pillow & 9.4.0 & GHSA-3f63-hfp8-52jq & CVE-2023-50447 \\
removed & PyPI & pillow & 9.5.0 & GHSA-j7hp-h8jx-5ppr & CVE-2023-4863 \\
removed & PyPI & pillow & 9.5.0 & PYSEC-2023-175 & -- \\
removed & PyPI & pillow & 9.5.0 & PYSEC-2023-227 & CVE-2023-44271 \\

\multicolumn{6}{l}{\textit{poetry}} \\
added & PyPI & poetry & 2.3.1 & GHSA-2599-h6xx-hpxp & CVE-2026-34591 \\
added & PyPI & poetry & 2.3.2 & GHSA-2599-h6xx-hpxp & CVE-2026-34591 \\

\multicolumn{6}{l}{\textit{tornado}} \\
added & PyPI & tornado & 6.5.4 & GHSA-fqwm-6jpj-5wxc & CVE-2026-35536 \\
added & PyPI & tornado & 6.5b1 & GHSA-fqwm-6jpj-5wxc & CVE-2026-35536 \\

\multicolumn{6}{l}{\textit{werkzeug}} \\
removed & PyPI & werkzeug & 3.1.5 & GHSA-29vq-vm6x & CVE-2026-27199 \\

\bottomrule
\end{tabular*}

\par\vspace{0.6em}
\raggedright
\footnotesize
\textit{Note:} ``Removed'' and ``Added'' refer only to differences between
the two generated ground-truth snapshots, not to deletion from or insertion
into OSV or other upstream vulnerability databases.
\end{table*}

\end{document}